\documentclass[superscriptaddress,twocolumn,bibnotes,
amsmath,amssymb,aps,pra,floatfix]{revtex4-2} 

\usepackage{graphicx}
\usepackage{dcolumn}
\usepackage{bm}
\usepackage{hyperref}
\usepackage{physics}
\newcommand{\RomanNumeralCaps}[1]
    {\MakeUppercase{\romannumeral #1}}
\usepackage{braket}
\usepackage{siunitx} 
\usepackage{multirow} 
\usepackage{xcolor}

\newcommand{\kq}{k_\mathrm{q}}
\newcommand{\be}{\begin{equation}} 
\newcommand{\ee}{\end{equation}} 
\newcommand{\jhat}{\hat\jmath}

\begin{document}

\title{\texorpdfstring{Probing open- and closed-channel p-wave resonances}{}}

\author{Denise J.\ M.\ Ahmed-Braun}
\affiliation{Department of Physics, Eindhoven University of Technology, The Netherlands}
\author{Kenneth G.\ Jackson}
\affiliation{Department of Physics, University of Toronto, Canada}
\author{Scott Smale}
\affiliation{Department of Physics, University of Toronto, Canada}
\author{Colin J.\ Dale}
\affiliation{Department of Physics, University of Toronto, Canada}
\author{Ben A.\ Olsen}
\affiliation{Yale-NUS College, Singapore}
\author{Servaas J.\ J.\ M.\ F.\ Kokkelmans}
\affiliation{Department of Physics, Eindhoven University of Technology, The Netherlands}
\author{Paul S.\ Julienne}
\affiliation{Joint Quantum Institute, NIST and the University of Maryland, U.S.A.}
\author{Joseph H.\ Thywissen}
\affiliation{Department of Physics, University of Toronto, Canada}

\date{\today}

\begin{abstract} 
We study the near-threshold molecular and collisional physics of a strong \textsuperscript{40}K p-wave Feshbach resonance through a combination of measurements, numerical calculations, and modeling. Dimer spectroscopy employs both radio-frequency spin-flip association in the MHz band and resonant association in the kHz band. Systematic uncertainty in the measured binding energy is reduced by a model that includes both the Franck-Condon overlap amplitude and inhomogeneous broadening. Coupled-channels calculations based on mass-scaled \textsuperscript{39}K potentials compare well to the observed binding energies and also reveal a low-energy p-wave shape resonance in the open channel. Contrary to conventional expectation, we observe a nonlinear variation of the binding energy with magnetic field, and explain how this arises from the interplay of the closed-channel ramping state with the near-threshold shape resonance in the open channel. We develop an analytic two-channel model that includes both resonances as well as the dipole-dipole interactions which, we show, become important at low energy. Using this parameterization of the energy dependence of the scattering phase, we can classify the studied \textsuperscript{40}K resonance as broad. Throughout the paper, we compare to the well understood s-wave case, and discuss the significant role played by van der Waals physics. The resulting understanding of the dimer physics of p-wave resonances provides a solid foundation for future exploration of few- and many-body orbital physics. 

\end{abstract}
\maketitle

\section{Introduction \label{sec:intro}}

Strong p-wave interactions are rare in nature, so their extreme tunability in ultracold systems \cite{Regal:2003go,Zhang:2004cy} is an opportunity for discovery \cite{Gurarie:2005it,Cheng:2005kv,Iskin:2006ep}. 
Despite recent advances in understanding, such as universal relations for p-wave systems \cite{Yoshida:2015hh,Yu:2015go,Luciuk:2016gr,He:2016bn,HuiHu:2016}, 
open questions remain, including the effect of confinement on Feshbach dimers \cite{Granger:2004im,Kanjilal:2004jv,Cui:2012be,Hess:2014hd,Gao:2015hz,Saeidian:2015db,Wang:2016ec,Waseem:2016ki,Zhang:2017fh,Waseem:2017ep,Fonta:2020ev,Chang:2020,Marcum:2020,Fonta:2020ev} and correlation strength \cite{Cui:2016kd,He:2016gh,Yin:2018hl}. One-dimensional systems hold the prospect for duality between strongly interacting odd waves and weakly interacting even waves  \cite{Cheon:1999jj,Girardeau:2005dt,Girardeau:2006dz,Sekino:2018gk}, for a topological phase transition in two-dimensional systems \cite{Read:2000iqa,Tewari:2007fj}, and for engineered states \cite{Jiang:2016em,Yang:2015ul,Hu:2016di}. Even in three-dimensional systems, p-wave trimer states have yet to be observed \cite{JonaLasinio:2008il,Braaten:2012go,Pricoupenko:2018ge,Pricoupenko:2019bt}.

Experimental work on ultracold p-wave alkali systems has focused on the fermionic isotopes $^{40}$K \cite{Regal:2003go,Gunter:2005er,Gaebler:2007} and $^6$Li \cite{Zhang:2004cy,Chevy:2005im,Schunck:2005cf,Fuchs:2008ka,Nakasuji:2013gw,Waseem:2017ep,top2020spinpolarized}, in part because s-wave collisions are easily suppressed with spin polarization. Experimental investigations have included studies of elastic and inelastic collision rates \cite{Chevy:2005im,Inada:2008hz,Nakasuji:2013gw,Waseem:2017ep,top2020spinpolarized}, spectroscopy \cite{Gaebler:2007,Fuchs:2008ka,Luciuk:2016gr}, and low-dimensional confinement \cite{Gunter:2005er,Chang:2020,Waseem:2016ki,Marcum:2020}. 

In this work, we perform association spectroscopy to determine the binding energies of p-wave Feshbach dimers near a strong resonance of $^{40}$K. To explain these measurements, we offer a new analytic treatment that builds on the commonly used effective-range approximation (ERA) of p-wave scattering \cite{Bethe1949,OMalley:1961gh,Gao:1998gi,Zhang:2010gc,Crubellier:2019kw},
\be \label{eq:effrange}
\cot \delta = -(V k^3)^{-1} - (R k)^{-1} + \mathcal{O}\{k\}, 
\ee
where $\delta$ is the scattering phase, $\hbar k$ is the relative momentum, $V$ is the ($k$-independent) scattering volume, and $R>0$ is the effective range. An alternate formulation is the unitary $S$-matrix element, $S = \exp(2 i \delta)$, such that
\be \label{eq:Seffrange}
S = \frac{V^{-1} + R^{-1} k^2 + \mathcal{O}\{k^4\} -i k^3 - \mathcal{O}\{i k^7\}}{V^{-1} + R^{-1} k^2 + \mathcal{O}\{k^4\} +i k^3 + \mathcal{O}\{i k^7\}}
\ee
is an equivalent approximation \footnote{For real momenta $k$, $\delta$ is a real and odd function of $k$, so that $k^3 \cot\delta$ has only even powers of $k$. As written in Eq.~\eqref{eq:Seffrange}, $S=(-k^3 \cot\delta - i k^3)/(-k^3\cot\delta + i k^3)$, so that the real parts of the numerator and denominator have only even orders, and the imaginary part is odd in $k$. Beyond the ERA, the $S$-matrix may of course take other forms. However when finding the low-energy values of $V$ and $R$, the $k^2$ correction to the leading imaginary term must be divided out, so that $R$ captures all of the quadratic-in-$k$ correction to the phase shift. For this reason, after $ik^3$, the first neglected term in imaginary part of Eq.~\eqref{eq:Seffrange} must be $\mathcal{O}\{ i k^7 \}$}. 

However, the ERA is invalidated at low energy due to a divergent contribution to the scattering volume from the weak $1/r^3$ dipole-dipole potential. We offer a more complete parameterization of terms in the scattering phase shift by factoring the $S$-matrix into three terms: 
$S_\mathrm{dip}$ for dipole-dipole interactions, $S_\mathrm{P}$ for the entrance channel, and
$S_\mathrm{FB}$ for the Feshbach mechanism. In the $^{40}$K case $S_\mathrm{P}$ has a shape resonance
and causes the ERA to become inaccurate for the largest binding energies we measure. Nonetheless, $V$ and $R$ provide a useful reference, since the ERA is appropriate for intermediate energies, and the correct low-energy limit for $S_\mathrm{P} S_\mathrm{FB}$. 

The Feshbach resonance \cite{Chin2010} tunes the scattering phase primarily through the scattering volume, conventionally written as 
\be \label{eq:scattvol} 
V(B) =  V_\mathrm{bg} \left(1 - \frac{\Delta}{\delta B} \right) \ee
where $V_{\rm bg}$ is the background scattering volume, $\delta B = B-B_{0}$, $B$ is the magnetic field, $B_0$ is the location of the resonance, and $\Delta$ is its magnetic width. We explain how this form emerges from the low-energy limit of a two-channel model in the broad- and narrow-resonance cases. We also discuss how the $B$-field variation of $R$ is coupled to $V(B)$ and linked to both $V_\mathrm{bg}$ and the van der Waals volume.

Just below resonance ($V > 0$), scattering is controlled by a low-energy bound-state pole of $S$ (where $\cot \delta = i$) located at $k=i\kappa$, with $\kappa>0$ given by 
\be \label{eq:kappaERA}
\kappa = \sqrt{\frac{3 R}{4 V}} \sec\left[ \frac{1}{3} \cos^{-1}\left( \frac{-3^{3/2} x}{2} \right) \right],
\ee
where $x \equiv (R^3/|V|)^{1/2}$ is a dimensionless parameter. The bound-state energy is $E = -E_\mathrm{b} = -\hbar^2 \kappa^2/m$, where $m$ is the atomic mass of $^{40}$K. Its series expansion for $x \ll 1$ is
\be \label{eq:EbERA}
E_\mathrm{b} \approx \frac{\hbar^2 R}{m V} \left( 1 + x + \frac{3}{2} x^2 + \frac{21}{8} x^3 + \ldots \right). \ee
In contrast to an s-wave Feshbach resonance, where the dimer binding energy curves quadratically with magnetic field towards the collision threshold, the p-wave $E_\mathrm{b}$ tends to scale linearly across threshold, as the Feshbach dimer state is confined by the centrifugal barrier. One can see this from the near-linearity of $V^{-1}$ in Eq.~\eqref{eq:scattvol}, for $|\delta B| \ll \Delta$. Also in contrast to the s-wave case, the binding energy depends on the effective range to lowest order. 

At the other side of the Feshbach resonance ($V < 0$), scattering is controlled by a pole in the fourth quadrant of the complex $k$ plane, adding a width to the resonance 
\footnote{The pole is located at 
$$k_\mathrm{pole} = \sqrt{\frac{3 R}{4 |V|}} \sec\left[ \frac{1}{3} \cos^{-1}\left( \frac{3^{3/2} i x}{2} \right) \right].$$ This corresponds to a pole in the ``non-physical'' Riemann sheet of $f(E)$, at 
$\Re E_\mathrm{pole} \approx E_0 \left( 1 - 3x^2/2 + \ldots \right)$
and 
$\Im E_\mathrm{pole} \approx -({\Gamma_0}/{2}) \left( x - 21 x^3/8 + \ldots \right)$.}. 
Although this pole controls the scattering phase, it does not correspond to a true molecular state. Instead, it creates a positive-energy scattering resonance ($\cot \delta = 0$) at 
\be \label{eq:E0ERA}
E_0 = -\frac{\hbar^2}{m}\frac{R}{V} \ee 
with which the resonant contribution to the phase shift can be written $\cot \delta(E) = -E_R^{1/2}(E - E_0)E^{-3/2}$, where $E_R = \hbar^2 m^{-1} R^{-2}$. An approximate form of the angle-averaged scattering cross section, $4 \pi k^{-2} \sin^2 \delta$ in each $M_L$ channel, is a Lorentzian with half-width 
\be \label{eq:Gamma0}
\frac{\Gamma_0}{2} = \frac{E_0^{3/2}}{E_R^{1/2}} \, .\ee 
We see that unlike s-wave scattering, ultracold p-wave scattering is energetically narrow: $\Gamma_0/E_0 \to 0$ as $E_0 \to 0$. For this reason, the near-threshold resonance is commonly referred to as a ``quasi-bound'' state. The nature of these states is further illustrated in Sec.~\ref{sec:ccc}. 

The following sections explore the near-threshold molecular and collisional physics of the commonly used p-wave resonance of $^{40}$K around 198.5\,G. In Sec.~\ref{sec:spectroscopy} we determine the dimer resonance locations as a function of magnetic field  using analytic models for the lineshapes. These measurements extend the pioneering work of Gaebler {\it et al.}~\cite{Gaebler:2007} to higher precision and to a wider range of magnetic fields. Energies are compared to a full coupled-channels calculation (Sec.~\ref{sec:ccc}) that updates prior work~\cite{Ticknor:2004}, enables us to identify the molecular physics that creates the Feshbach resonance, and allows us to identify the range of validity of simplified models. 
We find a departure from linear variation of the p-wave binding energy versus magnetic field and explain its origin in the coupling to a shape resonance above threshold. In Sec.~\ref{sec:twochannel}, we develop an analytic two-channel model that treats both resonances in the open and closed channels. Here, strong coupling manifests as a field dependence of the effective coupling between the channels. 
In Sec.~\ref{sec:parameterization} we provide a new parameterization of p-wave scattering based on this model. We summarize in Sec.~\ref{sec:conclusion}. 

\begin{figure*}[htb]
\centering
\includegraphics[width=7in]{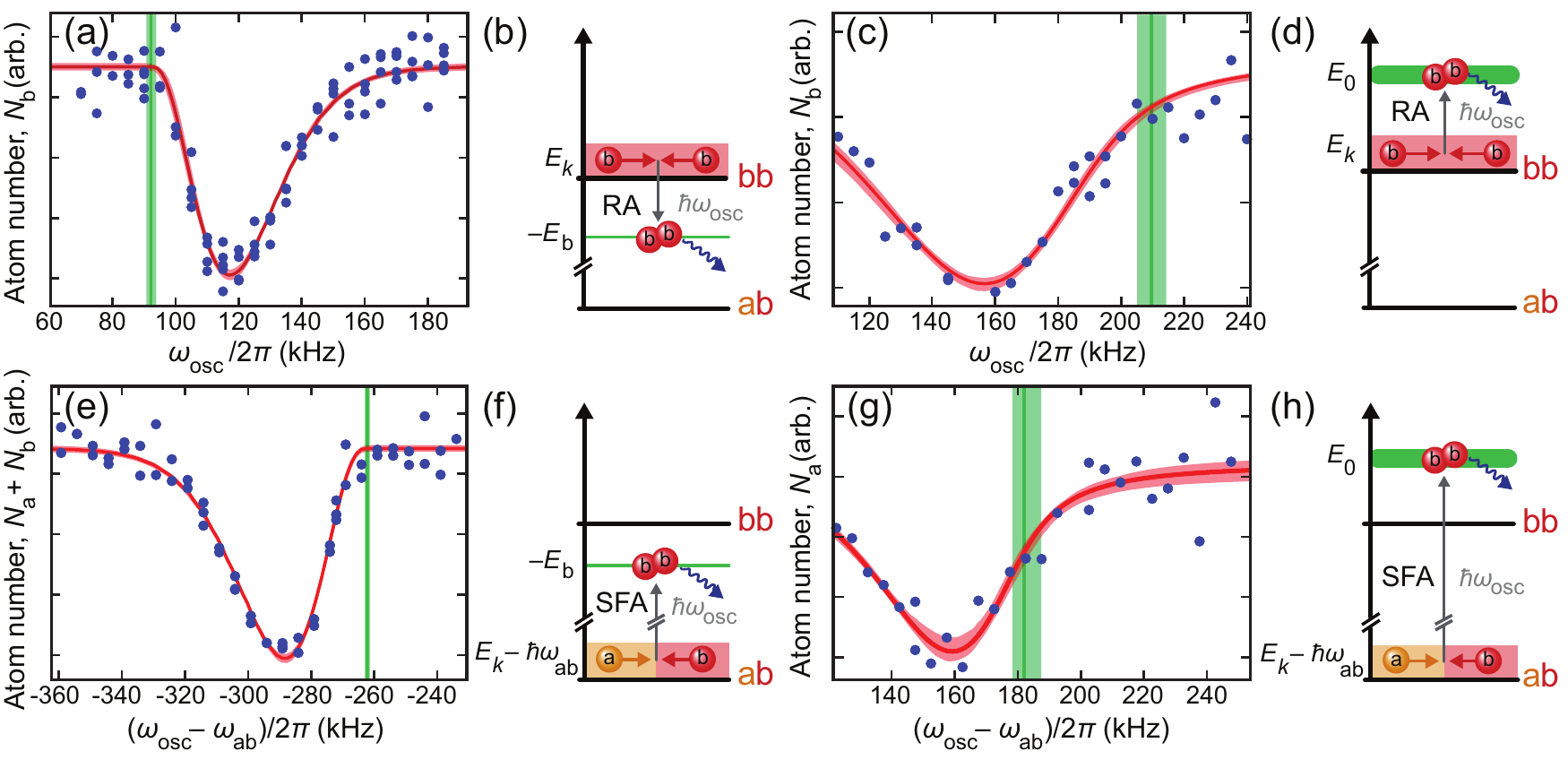}
\caption{{\bf Dimer association spectroscopy.} (a, b) Resonant association to a bound state: atom number remaining versus RA oscillation frequency at $197.80(1)$\,G. The observed resonance is a $|M_L|=1$ bound state with $E_\mathrm{b}/h = 92.9(1.5)$\,kHz, indicated by a green vertical line. 
(c, d) Resonant association to a quasi-bound state at $199.30(1)$\,G, corresponding to  $E_0/h = 210(4)$\,kHz in the $|M_L|=1$ channel. 
(e, f) Spin-flip association to a bound state: atom number remaining versus the difference between SFA oscillation frequency and the spin-flip frequency at $197.00(1)$\,G. The observed resonance is a $|M_L|=1$ bound state with $E_\mathrm{b}/h = 262.7(8)$\,kHz. 
(g, h) Spin-flip association to a quasi-bound state at $199.30(1)$\,G, corresponding to a quasi-bound state in the $|M_L|=1$ channel with energy $E_0/h = 182(5)$\,kHz. 
In (a, c, e, g) the solid red lines show fits of the spectroscopic data with Eqs.~(\ref{eq:AtomNumberBound1}), (\ref{eq:AtomNumberQuasiBound}), and (\ref{eq:AtomNumberQuasiBoundRA}). 
The shaded bands around the lineshapes (in red) and resonance locations (in green) represent 1$\sigma$ confidence intervals. 
Insets (b, d, f, h) illustrate each spectroscopy protocol. 
Dimer decay (represented by the blue arrow) produces untrapped final states and the loss signature. 
\label{fig:Spectra} }
\end{figure*}

\section{Dimer Spectroscopy \label{sec:spectroscopy}}

Fermionic $^{40}$K has a nuclear spin of four and a $^2$S$_\frac12$ ground state, giving rise to two hyperfine states in the ground state manifold with total spins $f=9/2$ and $7/2$, respectively having ten and eight spin components with projection $m_f$ 
\footnote{Although $f$ is not a good quantum number for finite $B$, it nevertheless remains a good approximate quantum number for labeling purposes, becoming exact at zero field.}. %
We use the convention of labeling these states $\ket{a}$, $\ket{b}$, $\ket{c}$, $\ldots$ in order of increasing energy at nonzero magnetic field. Due to the inverted hyperfine structure of the $^{40}$K atom, the lowest energy state $\ket{a}$ is $|f,m_f \rangle = |9/2,-9/2\rangle$, the next state $\ket{b}$ is $|9/2,-7/2\rangle$, and so forth. The p-wave ($L=1$) Feshbach resonances of interest here live in the $\ket{bb}$ entrance channel near 198.3\,G and 198.8\,G. 

\subsection{Association spectra}
\label{ssec:Association Spectra}

Dimer spectroscopy typically begins with a gas of $3 \times 10^4$ atoms held in a crossed optical dipole trap with a mean trap frequency of $2 \pi \times 320$\,Hz, at temperatures ranging between $0.2$\,$\mu$K and $0.5$\,$\mu$K. Microfabricated silver wires on an atom chip several hundred microns from the atomic cloud create oscillating magnetic fields that drive molecular association. Since the average collisional energy is comparable to $10$\,kHz, the order-$100$-kHz dipolar splitting between the $M_L=0$ and $|M_L|=1$ scattering channels is well resolved, which is an advantage for the study of p-waves over $^6$Li, for which $|M_L|$ channels are an order of magnitude closer~\cite{Zhang:2004cy,Chevy:2005im,Fuchs:2008ka,Waseem:2016ki,Gerken:2019cn}. In neither species can the splitting between $M_L=+1$ and $-1$ due to spin-orbit coupling~\cite{Bing:2019} be resolved. Our magnetic field, calibrated via the Breit-Rabi formula, has a statistical uncertainty of 10\,mG due to a combination of magnetic field noise and long-term drift.

Dimer binding energies are measured in two different protocols: resonant association (RA) and spin-flip association (SFA). For RA, also used in the first observations of p-wave dimer states  \cite{Gaebler:2007,Fuchs:2008ka}, a spin-polarized cloud is prepared in the $\ket{b}$ state, and $B$ is tuned to the desired value, between 195\,G and 200\,G. The oscillating field direction is aligned with the Feshbach field, driving $\Delta m_f=0$ transitions from free pairs of atoms to dimers, as illustrated in Figs.~\ref{fig:Spectra}(b) and \ref{fig:Spectra}(d). Since the initial and final states share the same continuum threshold, free atoms with energy $E_k$ are able to associate into either bound dimers with energy $E_k - \hbar \omega_\mathrm{osc}$, or quasi-bound dimers with energy $E_k + \hbar \omega_\mathrm{osc}$, where $\omega_\mathrm{osc}$ is the oscillation frequency of the field. 
The cloud is released from the trap and imaged after Stern-Gerlach separation to count atoms remaining in each spin state. Typical frequency-dependent loss curves are shown in Figs.~\ref{fig:Spectra}(a) and~\ref{fig:Spectra}(c). Atom number, imaged here at 209\,G, is a signature of molecular association since dimers decay on a millisecond time scale, through several mechanisms~\cite{Regal:2003go,Chevy:2005im,Waseem:2017ep}. At low density, loss is due to dipolar relaxation~\cite{Ticknor:2004,Chevy:2005im} to the open $\ket{ab}$ channel (see Sec.~\ref{sec:ccc} and Fig.~\ref{fig:DecayRates}), whose release energy ejects the pair from the optical trap (of depth $\lesssim 0.1$\,MHz) with high efficiency. At high density, three-body loss becomes increasingly important~\cite{Regal:2003go,Chevy:2005im}. Our analysis assumes the combined loss rate of dimers is fast, so that it is a faithful signature of association. 

The SFA protocol begins with a mixture of free $\ket{a}$ and $\ket{b}$ atoms. Spin-flip transitions between these states are induced by the $\sigma^+$-polarization component of the radio-frequency (rf) field near $44$\,MHz, as illustrated in Figs.~\ref{fig:Spectra}(f) and \ref{fig:Spectra}(h). Typical SFA spectroscopic curves are shown in Figs.~\ref{fig:Spectra}(e) and~\ref{fig:Spectra}(g).

Comparing Fig.~\ref{fig:Spectra}(e) to~\ref{fig:Spectra}(a), we see that the asymmetry of bound spectra inverts, since for SFA the dimer energy is always $E_k + \hbar \omega_\mathrm{osc} - \hbar \omega_{ab}$, where $\omega_{ab}$ is the bare spin-flip transition frequency, whereas for RA the dimer energy is $E_k-\hbar \omega_\mathrm{osc}$ for bound states.

The loss signatures shown in Figs.~\ref{fig:Spectra}(c) and~\ref{fig:Spectra}(g) differ, for the following reason. 
In the SFA protocol, the creation of a quasi-bound dimer is tagged by the conversion of an $\ket{a}$ to a $\ket{b}$ atom, which is not reversed by dissociation into a pair of $\ket{b}$ atoms. 
In the RA protocol, quasi-bound dimers that decay through the centrifugal barrier are not necessarily lost, but can simply re-convert to two free atoms. We correct for this in the lineshape model, as presented in Sec.~\ref{sssec:Free-to-quasi bound transitions}. 
Despite these differences, the dimer energies determined by these two protocols agree within experimental uncertainty. 

\subsection{Lineshapes and atom loss \label{ssec:lineshape}} 
In order to fit the spectral lines, we start with an analysis of the transition rate $\gamma$ from an initial free state $\ket{\mathrm{i}}$ to a final (quasi-)bound state $\ket{\mathrm{f}}$. We assume the role of atom-dimer coherence is negligible, since the pulses are long compared to the dimer lifetime. (This is supported by calculations of the dimer lifetime, shown in Fig.~\ref{fig:DecayRates} and discussed in Sec.~\ref{sec:ccc}.)

To first order, the transition rate is 
\begin{align}
\label{eq:FermisGoldenRule}
\gamma = \frac{2\pi}{\hbar} \abs{\braket{\mathrm{f}| H'|\mathrm{i} }}^2,
\end{align}
with the perturbing Hamiltonian $H'$ that drives the transition. 
This rate scales linearly with the Franck-Condon (FC) factor $F_{\mathrm{fi}}$, where
\begin{align}
\label{eq:FCgeneral}
F_{\mathrm{fi}} = \abs{\int \psi_\mathrm{i}^*(r) \psi_\mathrm{f}(r) dr}^2,
\end{align}
with energy-normalized incident wave function of relative motion $\psi_\mathrm{i}(r)$, with internuclear separation $r$, living in the entrance channel and (quasi-)bound-state wave function $\psi_\mathrm{f}(r)$ living in the outgoing channel. 
As outlined in Sec.~\ref{ssec:Association Spectra}, the RA protocol does not involve channel transitions.  In this situation both $\psi_\mathrm{i}(r)$ and $\psi_\mathrm{f}(r)$ correspond to the nonorthogonal wave function components in the entrance $\ket{bb}$ channel of the multichannel Feshbach system described in App.~\ref{sec:TransitionRateappendix}.

Knowing how to compute the lineshape for both protocols, we are now left with the task to relate the formation of (quasi-)bound states to the experimentally observed atom-loss. To ease this procedure, we note the analogy between the protocols used here and the photoassociation (PA) process as discussed in Ref.~\cite{Jones:JonesReview2006}. 
Using a perturbative approach and the approximations outlined in App.~\ref{sec:AtomLossappendix}, we find that in the case of free-to-bound-state transitions, the number of atoms lost from the trap is 
\begin{align}
\label{eq:AtomNumberBound1}
\delta N(\omega_\mathrm{osc}) = A_N P(T,E_k)F_{\mathrm{fi}},
\end{align}
where $A_N$ is an undetermined proportionality coefficient, $P(T,E_k)$ represents the thermal distribution of the incoming (free) particles, 
and $E_k = - E_\mathrm{b} + \hbar \omega_\mathrm{osc}$ in the RA case, or $E_k = -E_\mathrm{b} -\hbar \omega_\mathrm{osc} + \hbar \omega_{{ab}}$ in the SFA case, as discussed 
above (Sec.~\ref{ssec:Association Spectra}). 
We analyze our data with a Fermi-Dirac distributed $P(T,E_k)$, and compare to the use of a Maxwell-Boltzmann distribution in Sec.~\ref{ssec:Fitting and systematics}. 

In the case of free-to-quasi-bound transitions, we need to consider off-resonant transitions. For the SFA protocol, Eq.~\eqref{eq:AtomNumberBound1} is replaced with 
\begin{align}
\label{eq:AtomNumberQuasiBound}
\delta N^{\mathrm{SFA}} = A_N' \int^{\infty}_{k_\mathrm{min}} \!
P(T,E) F_{\mathrm{fi}}(k_{\mathrm{q}}) dk_{\mathrm{q}},
\end{align}
where $E = E_\mathrm{q}-\left(\hbar \omega_{\mathrm{osc}} -\hbar \omega_{{ab}}\right)$, $E_\mathrm{q} = \hbar^2 \kq^2/m$, and $k_\mathrm{min} = {\sqrt{{m (\omega_{\mathrm{osc}}-\omega_{{ab}})}/\hbar}}$. 
For the RA protocol, 
\be
\label{eq:AtomNumberQuasiBoundRA}
\delta N^{\mathrm{RA}} = A_N' \! \int^{\infty}_{k_\mathrm{min}} \!\! [P(T,E) 
- P(T,E_{\mathrm{q}})] F_{\mathrm{fi}}(k_{\mathrm{q}}) dk_{\mathrm{q}},
\ee
where $E = E_\mathrm{q}-\hbar \omega_{\mathrm{osc}}$, $k_\mathrm{min} = \sqrt{{m \omega_{\mathrm{osc}}}/{\hbar}}$ and where the second term on the right hand side of Eq. \eqref{eq:AtomNumberQuasiBoundRA} has been added to correct for the zero-transitions in the $\ket{bb}$ channel.
The implicitly assumed hierarchy of spectral widths is discussed in more detail in App.~\ref{sec:AtomLossappendix}.
Equations \eqref{eq:AtomNumberBound1} through \eqref{eq:AtomNumberQuasiBoundRA} indicate how the atom loss is directly related to the lineshape. The calculation of the lineshape in the case of free-to-bound and free-to-quasi-bound transitions will be discussed in the following two subsections. 

\subsubsection{Free-to-bound transitions \label{sssec:Free-to-bound transitions}} 
Starting our analysis of the Franck-Condon factor for free-to-bound transitions, we consider the overlap between the radial component of the bound-state wave function $\psi_\mathrm{f}(r) = u_{\mathrm{b}}(r)/r$ and the scattering state $ \psi_\mathrm{i}(r) = u_{\mathrm{sc}}(r)/r$. 
In the asymptotic region, $r \geq r_c$, with short-range cutoff $r_c$, 
the bound state is \cite{Yu:2015go}
\begin{align}
\label{eq:uBound}
u_{\mathrm{b}}(r) = \sqrt{r_c}\left(\frac{1}{r}+\kappa\right)e^{-\kappa r}\, .
\end{align}
The radial component of the energy-normalized scattering state in the asymptotic region in turn can be expressed as 
\begin{align}
\label{eq:uScatter}
u_{\mathrm{sc}}(r) = \sqrt{\frac{m}{\pi \hbar^2 k}} \left[\cos (\delta)\, \jhat_1(kr) + \sin(\delta) \, \hat n_1(kr) \right],
\end{align}
with $k$-dependent scattering phase $\delta$ and Ricatti-Bessel functions $\jhat_1$ and $\hat{n}_1$. 
Whereas the above asymptotic wave functions do not capture the correct behavior at short range, we approximate the rapid and out-of-phase oscillations of $u_{\mathrm{sc}}(r)$ and $u_{\mathrm{b}}(r)$ that occurs for deep potentials to lead to a vanishingly small overlap. Consequently, we neglect the short-range contribution to the FC overlap. \par 
Substitution of Eqs.~\eqref{eq:uBound} and~\eqref{eq:uScatter} into Eq.~\eqref{eq:FCgeneral} results in the overlap
\begin{align}
\label{eq:FCfreetoboundfull}
F_{\mathrm{fi}} = &\frac{m}{\pi \hbar^2} \frac{e^{-2 \kappa}}{k^3 r_c (k^2+\kappa^2)^2}  \Big|k r_c \kappa^2 \cos(k r_c +\delta)  \notag \\ & -\left[\kappa^2+k^2\left(1+r_c \kappa\right)\right]\sin(k r_c + \delta)\Big|^2.
\end{align}
In the experimentally relevant regime, we can simplify Eq.~\eqref{eq:FCfreetoboundfull} significantly. 
In our measurements, a typical collision energy is $5$\,kHz, and binding energies range between 14\,kHz and 700\,kHz. The maximum value of $R$ at the Feshbach resonance is $\approx 1.16 r_\mathrm{vdW}$~\cite{Gao:2009ek,Hammer:2010hm,Braaten:2012go}, where $r_\mathrm{vdW} \approx \SI{65.0223}{\bohr}$ for $^{40}$K \cite{Falke:2008dq}; for this resonance, $R\sim \SI{50}{\bohr}$, as we discuss later. Since $R$ and $r_\mathrm{vdW}$ are comparable, so are the energy scales that correspond to them, $E_R$ and $E_\mathrm{vdW}$, both on the 20\,MHz scale. (See further discussion in Sec.~\ref{ssec:vdW} and Sec.~\ref{sec:parameterization}.) 
We furthermore assume that the binding energy of the lowest-lying dimer in the open channel, $E_\mathrm{b}'$, is set only by van der Waals physics. In sum, the typical hierarchy of energy scales in our experiment is 
\begin{align}
 \label{eq:energies}
E_k , |E_\mathrm{b}|  \ll \{ E_R \sim E_\mathrm{vdW} \sim | E_\mathrm{b}' | \} \end{align}
Equivalently, there is a separation of length scales and momenta:
\begin{align}
 \label{eq:momenta}
 k \,\, \mbox{and} \,\, \{ \kappa \sim \sqrt{R/|V|} \} \quad \ll \quad \{ R^{-1} \sim r_\mathrm{vdW}^{-1} \}.
\end{align}
These inequalities give us two small parameters: 
\begin{align}
 \label{eq:smallparams}
k R \ll 1 \qquad \mbox{and} \qquad \sqrt{\frac{R^3}{|V|}} \ll 1 
\end{align}
where $\sqrt{{R^3}/{|V|}}$ was defined as $x$ in the Introduction. 

A natural scale for the cutoff $r_c$ is also set by the short-range length scale, $R \sim r_\mathrm{vdW}$. The asymptotic wave functions used have neglected the van der Waals $C_6/r^6$ potential, where $C_6 = 16 r_\mathrm{vdW}^4 \hbar^2/m$, but not the centrifugal $1/r^2$ barrier. The  interplay between these is indicated by the first zero-crossing of the combined effective potential, $-C_6 r^{-6}+(2\hbar^2/m)r^{-2}$, which is $2^{3/4} r_{\mathrm{vdw}}$. This is, for instance, the inner classical turning point of the centrifugal barrier in the low-energy limit. 
Since the resulting lineshape is independent of the precise cutoff chosen, and
$r_\mathrm{vdW}$ is comparable to $R$, we choose to fix $r_c=R$ from here forward.

Applying the small parameters to the computation of the overlap resulting from Eqs.~\eqref{eq:uBound} and~\eqref{eq:uScatter}, we find that the FC overlap for a free-to-bound transition can be approximated as
\begin{align}
\label{eq:FCbound}
F_{\mathrm{fi}} \approx \frac{m}{\pi \hbar^2} \frac{R k^3}{(k^2 + \kappa^2)^2} = \frac{1}{\pi} \frac{E_R^{-1/2} E_k^{3/2}}{(E_k + E_\mathrm{b})^2} 
\end{align}
In the high kinetic energy limit $E_k \gg E_\mathrm{b}$, Eq.~\eqref{eq:FCbound} scales as $E_k ^{-1/2}$, in agreement with the scaling law of the p-wave contact and with a numerical prefactor of 2$R$, corresponding to the value of the p-wave contact of a Feshbach dimer~\cite{Yoshida:2015hh,Yu:2015go,Luciuk:2016gr,He:2016bn,HuiHu:2016}. 
With Eq.~\eqref{eq:FCbound} and the thermal factor $P(T,E_k)$ discussed in Sec.~\ref{ssec:Fitting and systematics}, Eq.~\eqref{eq:AtomNumberBound1} can now be used to fit free-to-bound spectral data. 

\begin{figure*}[tb!]
\centering
\includegraphics[width=7in]{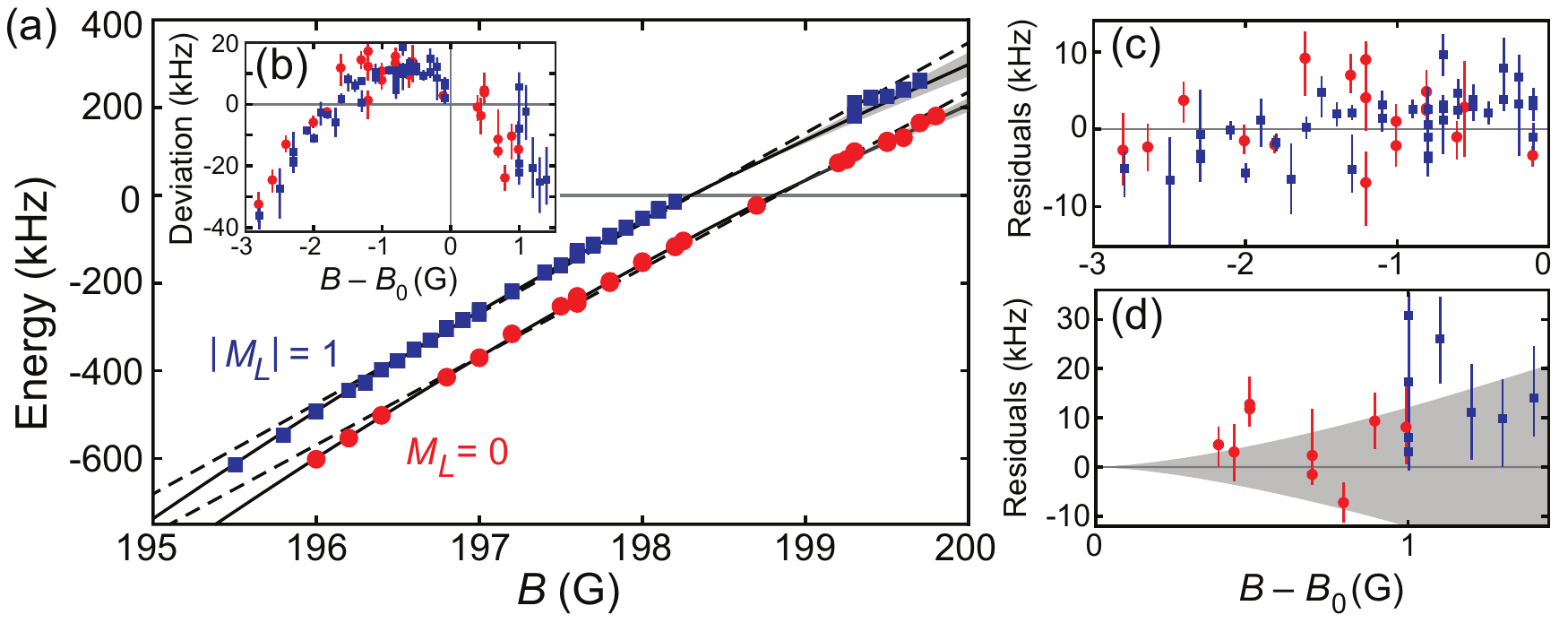}
\caption{{\bf p-wave dimer energy versus magnetic field.} 
(a) Experimental best-fit results for $M_L=0$ ($|M_L|= 1$) are shown in red (blue), with 1$\sigma$ error bars. The dashed lines show a linear fit,
and the solid lines show coupled-channels calculations of the dimer energy. The shaded region above resonance shows $\pm \frac12 \Gamma_0/h$, where $\Gamma_0$ is the calculated energetic width of the scattering resonance. 
(b) The deviation from a linear fit clearly shows the bending of the bound state near resonance as explained in Sec.~\ref{ssec:NearThreshold}. (c,d) The residual deviation between the data and the coupled-channels calculation. The shaded gray band gives the energetic width $\pm \frac12 \Gamma_0/h$ of the quasi-bound states. Since $\Gamma_0/2 \to E_0^{3/2} E_R^{-1/2}$ near threshold and the two components have similar $R$, the widths are indistinguishable on this plot. 
\label{fig:pwave}}
\end{figure*}

\subsubsection{Free-to-quasi-bound transitions \label{sssec:Free-to-quasi bound transitions}} 

Once the closed-channel dimer crosses threshold, it acquires a finite resonance width. As outlined in App. \ref{sec:AtomLossappendix}, this width has to be taken into account in the computation of the atom-loss, meaning that we need to consider the possibility of resonant as well as non-resonant transitions to the quasi-bound state. Additionally, we require the quasi-bound wave function $u_{\mathrm{q}}(r)$ to be of the form of the scattering wave function as presented in Eq.~\eqref{eq:uScatter}, as the wave function beyond the range of the potential barrier behaves as a scattered wave with wave number $\kq$ and phase shift $\delta_\mathrm{q}$. The phase shift of this scattered wave provides us with the scattering volume and the effective range of the $\ket{bb}$ channel, such that we can compute the energy of the positive-energy scattering resonance $E_0$ as detailed in Sec.~\ref{sec:intro}.

As described in App.~\ref{sec:FCappendix}, the implementation of Eq.~\eqref{eq:FCgeneral} allows for the formulation of an analytic expression of the FC overlap for a free-to-quasi-bound transition. In the case of the SFA protocol, we can additionally use the two small parameters as presented in Eq.~\eqref{eq:smallparams} in order to obtain the simplified expression
\begin{align}
\label{eq:FCQBapprox}
F_{\mathrm{fi}}(\kq) \approx \frac{m^2}{\pi^2 \hbar^4} \frac{ k^3 \sin^2(\delta_{\mathrm{q}})}{ k_{\mathrm{q}}^3\left(k^2-k_{\mathrm{q}}^2\right)^2} \qquad \text{for} \quad k \neq k_{\mathrm{q}}.
\end{align}
Analogously to the computation of the bound-state atom loss, we can straightforwardly substitute Eqs.~\eqref{eq:FCQBFull} and \eqref{eq:FCQBapprox} into Eqs.~\eqref{eq:AtomNumberQuasiBound} and \eqref{eq:AtomNumberQuasiBoundRA}, respectively, to fit the experimental atom loss and find the resonant energy $E_0$.

\subsection{Determination of the Feshbach dimer energy \label{ssec:Fitting and systematics}}

Dimer energies are determined by a fit of spectroscopic data to Eqs.~\eqref{eq:AtomNumberBound1}, \eqref{eq:AtomNumberQuasiBound}, and \eqref{eq:AtomNumberQuasiBoundRA}. 
We use the Monte Carlo Bootstrap method \cite{efron1979,bajorski} for fitting and uncertainty estimation. In brief, this method works as follows: we randomly sample $M$ times from the $M$ data points in each data set with replacement. 
The resulting collection of $M$ points (with individual data points randomly omitted or repeated) is fit, yielding best-fit parameters $E_\mathrm{b}$ or $E_0$, $N$, $T$, and $A_N$. This procedure is repeated 5000 times for bound-state data and 500 for quasi-bound data (due to constraints on computing time and the increased complexity of the quasi-bound fitting function). The strength of the method is that it does not rely upon a prior assumption of a probability distribution. Data sets are excluded when the distribution of best-fit parameters is significantly skewed or non-Gaussian. Otherwise, we take the median of the resulting parameter distributions as the overall best-fit parameters, and use 1$\sigma$ confidence intervals for uncertainties.

The vertical green bars in Fig.~\ref{fig:Spectra} show examples of binding energies determined by this procedure. It is striking, especially for Fig.~\ref{fig:Spectra}(a) and Fig.~\ref{fig:Spectra}(e), how far $E_\mathrm{b}/h$ is from the loss peak. The accuracy of the determination depends upon the FC factor, which adds a significant asymmetry to the thermally broadened lineshape. By comparison, we find that extrapolation of the frequency of maximum loss at finite $E_F$ to zero $E_F$, as used in Ref.~\cite{Gaebler:2007}, over-estimates the binding energy by roughly 10\,kHz in our typical experimental conditions. This emphasizes the critical role of the lineshape functions found in Sec.~\ref{ssec:lineshape}.

For these fits, we use a collisional factor based on the Fermi-Dirac distribution $\bar{n}(\mu,\epsilon)$, where $\mu$ is the chemical potential and $\epsilon$ is the single-particle energy. The probability distribution of relative momentum $k$ averaged over the inhomogeneous density distribution in the trap is
\be P_\mathrm{FD}(k) = \mathcal{N} k \int\!\! d^3 \bm{k}_\mathrm{av} \! \int\! d^3 \bm{r} \, \bar{n}(\mu_L,\epsilon_A) \, \bar{n}(\mu_L,\epsilon_B)
\ee
where $\bm{k}_\mathrm{av}$ is the average momentum, $\bm{r}=(\bm{x}\omega_x + \bm{y}\omega_y + \bm{z}\omega_z)/\bar\omega$ is the rescaled position in the trap, $\mu_L = \mu - m \bar\omega^2 r^2/2$ is the local chemical potential, and $\epsilon_{A,B} = (\hbar^2/2m)(\bm{k} \pm \bm{k}_\mathrm{av})^2$. Since the energy of the cloud is rotationally symmetric in free space, all three $M_L$ channels see the same distribution, and one can choose an arbitrary axis for $\bm{k}$ so long as $|\bm{k}|=k$. The leading factor of $k$ accounts for relative-velocity weighting of the the event rate \cite{Jones:JonesReview2006}. Here $\mathcal{N}$ is a normalization factor, chosen so that $\int P(T,E_k) dE_k=1$, where $P(T,E_k)dE_k = 4 \pi k^2 P_\mathrm{FD}(k) dk$. 
This treatment takes a semiclassical isotropic limit, which should be valid due to the large number of fermions in the trap, and relatively weak trap. 
We use $P_\mathrm{FD}$ distributions generated at the measured $T=0.3T_F$; re-fitting with $T=0.4T_F$ shifts binding energies by less than 1\,kHz. 
By comparison, a Maxwell-Boltzmann distribution was found to give a $\sim$3~kHz shift towards larger binding energy, roughly independent of magnetic field. We do not attempt to account for shifts in the distribution due to weak interactions, and in data acquisition avoid the strongly interacting regime ($-0.05$\,G$ < \delta B < 0.4$\,G, as delineated in Ref.~\cite{Luciuk:2016gr}).

Our trapping light introduces two possible systematics: AC Stark shifts and confinement-induced shifts. In the separated-atom limit, since our 1064-nm trapping beams are far-detuned compared to the hyperfine splitting, even circularly polarized trapping beams would create a negligible differential light shift. A coincidental molecular transition could cause a more significant shift, but none exists to our knowledge. 
Since the measured $E_\mathrm{b}/\hbar$ and $E_0/\hbar$ are much larger than the trap frequencies, confinement has a negligible effect on the dimer wave functions \cite{Zurn:2013}.
Similarly, the discretization of the continuum will affect the thermal model at the sub-kHz scale, smaller than statistical uncertainties. Our two-body FC coefficient does not take into account possible three-body processes \cite{Regal:2003go,Chevy:2005im,Nakajima:2011ic} or many-body correlations. We restrict our data collection to fields at which driven dimer association is expected to dominate atom loss.

Figure \ref{fig:pwave} shows the best-fit dimer energies from roughly eighty spectra, taken at magnetic fields ranging from 195.5\,G to 200\,G. 
A striking feature in our data is a nonlinearity in the binding energy versus magnetic field. Figure~\ref{fig:pwave}(b) shows the deviation from a linear trend, indicated by the dashed line in Fig.~\ref{fig:pwave}(a). Evidence for nonlinearity was first found in Ref.~\cite{Gaebler:2007}, since piecewise-linear fits to binding energy versus $\delta B$ above and below threshold did not lead to the same zero-energy Feshbach resonance location. Since our measurements probe a wider range of fields, the curvature in binding energy appears clearly. 
We find good agreement, especially for bound-state energies, with new coupled-channels calculations (shown as black lines and discussed in Sec.~\ref{sec:ccc}). 
Allowing for an overall magnetic-field shift in calculated binding energies, we find a best-fit $-7.3$\,mG with $1.0$\,mG statistical uncertainty and $5$\,mG estimated systematic uncertainty. Figure~\ref{fig:pwave}(c) and (d) show residuals of this comparison, with an rms scatter in $E_\mathrm{b}$ of 4\,kHz, comparable to the statistical errors of the individual spectra.

\begin{figure*}[tb!]
\centering
\includegraphics[width=7in]{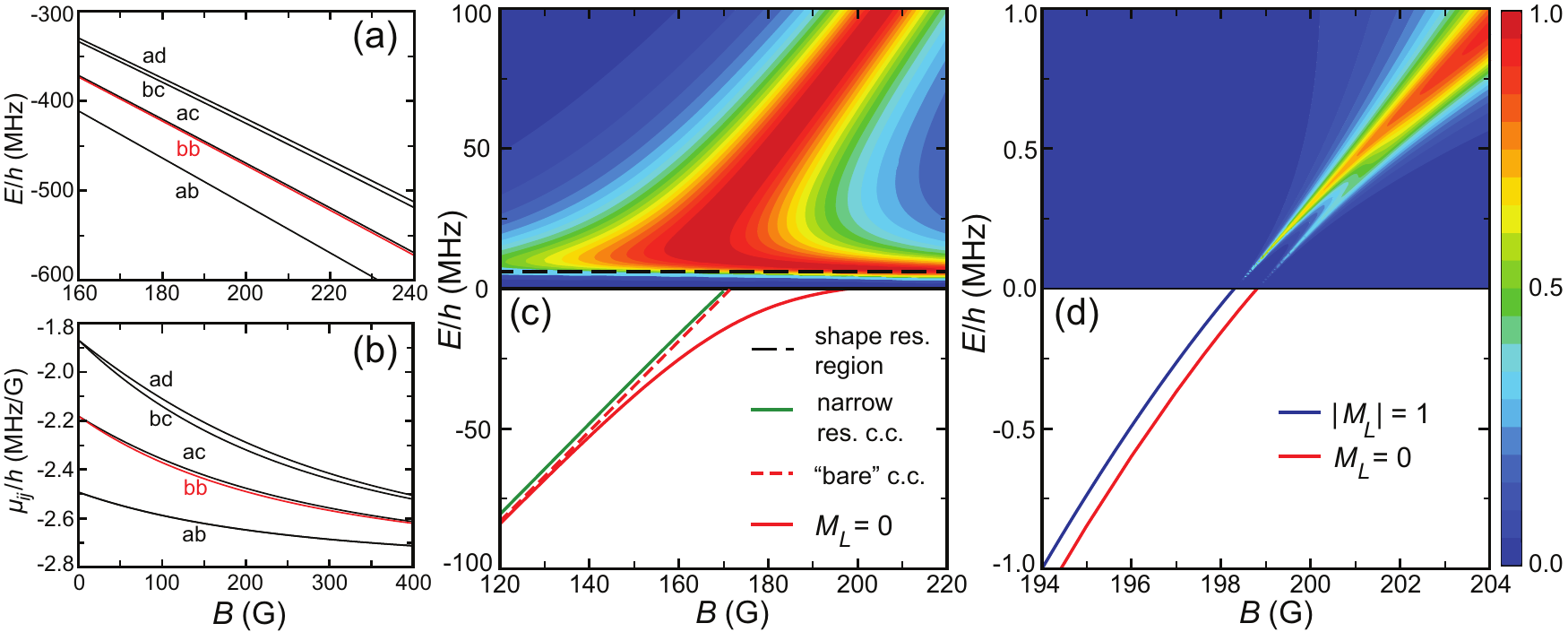}
\caption{{\bf p-wave properties versus magnetic field.} 
(a) Separated-atom energies of the five lowest spin channels of $f_1=9/2,f_2=9/2$ spin character in the p-wave Hamiltonian block for $M_\mathrm{tot}=-7$. 
(b) The magnetic moments of the five spin channels of lowest energy. 
(c) Calculated bound-state energies and scattering probabilities over a wide scale of $E$ and $B$ of $\pm100$\,MHz and 100\,G, where $E=0$ represents the separated atoms energy of the $\ket{bb}$ channel. The solid red line shows the $M_L=0$ level that makes the 198.8\,G Feshbach pole studied in the experiment; the dashed red line approximates the corresponding ``bare'' closed-channel state. The green line shows the closed-channel bound state that makes a narrow resonance near 170\,G (see App.~\ref{sec:NarrowFR}). The color contours above threshold show the loss $1-|S_{bb,bb}|^2$ from the $\ket{bb}$ channel; red indicates near unitary maximum loss and blue indicates minimal loss. The dashed black line indicates the region of the $\approx$ 7\,MHz shape resonance in the $\ket{bb}$ channel. Loss from $\ket{bb}$ is almost entirely due to strong decay of the closed-channel resonance to the $\ket{ac}$ channel. 
(d) Near-threshold properties over a scale of $\pm$1\,MHz and 10\,G. The blue and red lines show the CC bound-state energies of the $M_L=0$ (lower) and $|M_L|=1$ (upper) levels that cross threshold at 198.8\,G and 198.3\,G respectively. The color contours show the averaged near-threshold elastic scattering probabilities of the three $M_L$ components, Eq.~\eqref{eq:sum}. 
\label{fig:BigPicture}}
\end{figure*}

Figure \ref{fig:pwave}(d) shows increased scatter and a possible trend in the difference between measured and calculated $E_0$. 
This could be explained in part by heating and polarization of the cloud during spectroscopy. Quasi-bound dimers that decay through the centrifugal barrier create atoms with a relatively large kinetic energy, which rapidly heat the cloud. For SFA of quasi-bound atoms, a similar process also spin-polarizes the cloud since $\ket{a}$ atoms are irreversibly converted into high energy $\ket{b}$ atoms. We mitigate this effect by fitting just the remaining $\ket{a}$ atoms after our rf pulse [see Fig.~\ref{fig:Spectra}(g)]. 
Another systematic may come from increasing overlap between the $M_L=0$ and $|M_L|=1$ channels, which restricts the spectroscopic range. 
The linewidth of the scattering resonances increases roughly as $(B-B_0)^{3/2}$, causing the features to overlap beyond $\sim 201$\,G, as illustrated in Fig.~\ref{fig:BigPicture}(d) and discussed in the next section.

\section{Coupled-channels calculations \label{sec:ccc} }

We carry out standard coupled-channels (CC) calculations~\cite{Stoof1988,Hutson2008,Berninger2013} based on the known atomic matrix elements of the full spin Hamiltonian~\cite{Arimondo1977} and the $^1\Sigma_g^+$ (singlet) and $^3\Sigma_u^+$ (triplet) molecular potentials for the $^{40}$K$_2$ dimer molecule. We mass-scale the $^{39}$K$_2$ singlet and triplet potentials of Falke {\it et al.}~\cite{Falke:2008dq} without Born-Oppenheimer corrections and use the effective spin-dipolar coupling determined by Ref.~\cite{Xie2020}. We use the molscat~\cite{Hutson2019b,molscatNote} and bound~\cite{Hutson2019a,boundNote} packages to calculate the needed scattering $S$-matrix and near-threshold bound-state energies for two $^{40}$K atoms, as illustrated in Fig.~\ref{fig:BigPicture}. These are calculated without adjustable parameters. 
The excellent agreement between the CC predictions and the measured p-wave resonance is much better than the same comparison for the s-wave resonance (see App.~\ref{sec:swave}). This agreement in the p-wave case
 gives us confidence in the accuracy of CC predictions over wider range of field and energy near threshold. 

A scattering channel $f_1 m_{f_1} f_2 m_{f_2} L M_L$ is specified by the $f,m_f$ values of the two atoms that are interacting and the ``partial wave'' quantum numbers $L,M_L$ of their relative angular momentum. The total angular momentum projection quantum number 
\begin{equation}
M_\mathrm{tot} = m_{f_1}+m_{f_2}+M_L
\label{Mtot}
\end{equation}
is conserved, allowing us to block the Hamiltonian according to the $M_\mathrm{tot}$ value. 

Since we are interested in collisions of two $\ket{b}$ atoms, and identical fermions only collide with odd $L$ values, the threshold channels of interest are the $\ket{bb}$ p-wave ones with $L=1$ and $M_L = -1$, 0, or 1, for which $M_\mathrm{tot}=$ $-8$, $-7$, and $-6$ respectively.  These three $M_\mathrm{tot}$ values in turn give rise to Hamiltonian blocks with 8, 13, and 20 p-wave spin channels for the separated atom states $|ij,M_L\rangle$, where $i$ and $j$ represent the states ($a$, $b$, $c$, $\ldots$) consistent with Eq.~\eqref{Mtot} for a given $M_L$. Figures \ref{fig:BigPicture}(a) and \ref{fig:BigPicture}(b) show the channel energies and magnetic moments of the five lowest channels of the $M_\mathrm{tot}=-7$ block that contains the $\ket{bb,0}$ channel. The only open p-wave channels for low collision energy $E/h$ less than 1\,MHz in the 200\,G region are $|aa,+1\rangle$, $|ab,0\rangle$ and $|bb,-1\rangle$ for $M_\mathrm{tot}=-8$, $|ab,+1\rangle$ and $|bb,0\rangle$ for $M_\mathrm{tot}=-7$, and $|bb,+1\rangle$ for $M_\mathrm{tot}=-6$; all other channels are closed at such low energy (meaning the channel energy is larger than $E$).  

Bound states below the $\ket{bb}$ threshold with $M_\mathrm{tot}=-7$ and $-8$ can decay by spin-dipolar relaxation if they lie energetically above one of the p-wave open channels for some field $B$; however, bound states with $M_\mathrm{tot}=-6$ can  decay only to an $L=3$ f-wave open channel
\footnote{From the multichannel perspective, all of these states are quasi-bound. Here we choose to call them bound states if they are below threshold, in order to maintain the single-channel terminology.}.
We include f-wave basis functions in the CC calculations for the decay rates, but they are not necessary for the energy positions, which change negligibly (less than 1 kHz) when f-waves are introduced.   For simplicity in the following, when $M_\mathrm{tot}$ is specified, we suppress the implied $M_L$ in the ket notation.

\subsection{van der Waals character of threshold scattering \label{ssec:vdW} }

The CC calculations reveal the spin character of the bound and quasi-bound states of the $^{40}$K$_2$ dimer.  Due to the relatively low mass of the $^{40}$K atom, these states are sparse near threshold.  Furthermore, they are relatively easy to understand, although complicated by the spin mixing among the various spin channels.  It is easiest first to sketch out the vibrational and rotational character of these states in terms of the long-range van der Waals character of the long-range singlet and triplet potential with a leading term that varies as $-C_6/r^6$.  This potential is characterized by the length \cite{Gribakin1993,Gao1998a,Gao:1998gi} 
\be \label{abar}
\bar{a} = \frac{2\pi}{\Gamma(1/4)^2} \left ( \frac{2 m_r C_6}{\hbar^2} \right )^\frac{1}{4} = \frac{4\pi}{\Gamma(1/4)^2} r_\mathrm{vdW} \ee
and corresponding energy
\be \bar{E} = \frac{\hbar^2}{2m_r \bar{a}^2} \label{Ebar} \,, \ee
where $m_r$ is the reduced mass, such that $\bar{a} \approx 0.9559776\,r_\mathrm{vdW}$.  
Since both atoms are in electronic $^{2}\mathrm{S}_{1/2}$ ground states, the coefficient $C_6 \approx 3925.91$\,$E_\mathrm{h}a_0^6$~\cite{Falke:2008dq} is the same for all channels, where $E_\mathrm{h}$ is the Hartree energy, and $a_0$ is the Bohr radius. For two $^{40}$K atoms, $\bar{a}= 62.160\,a_0$ and $\bar{E}/h=23.375$\,MHz.

The spectrum of a van der Waals potential gives much insight into the states near threshold~\cite{Julienne2009b,Chin2010}. Quantum defect theory shows that given the s-wave scattering length $a$ of a van der Waals potential, the states and scattering properties near threshold of the other partial waves are also determined \cite{Gao:1998gi,Gao2001,Gao:2009ek}.  
When $a \gg \bar{a}$ the binding energy of the last s-wave bound state is universal, $E_{-1}=\bar{E}/(a/\bar{a})^2$, and the p-wave phase shift is~\cite{Idziaszek2010a}
\be \label{eq:eta1}
\tan\delta_{L=1}(k) \approx 2 k^3 \bar{V} \frac{a-\bar{a}}{a-2\bar{a}} \ee
where $\bar{V}$ is the van der Waals volume,
\be \label{eq:Vbar}
\bar{V} = \frac{\Gamma(1/4)^6}{144\pi^2\Gamma(3/4)^2}\bar{a}^3 =  \frac{4\pi}{9 \Gamma(3/4)^2} r_\mathrm{vdW}^3 .  \ee 
For two $^{40}$K atoms, $\bar{V} \approx (0.9760\,r_\mathrm{vdW})^3 \approx (63.464\,a_0)^3$. 

One sees from Eq.~\eqref{eq:eta1} that $ \tan\delta_{L=1}$ diverges at threshold when $a=2\bar{a}$.  This implies there is a p-wave bound state at $E=0$.  For a range of $a \gtrsim 2\bar{a}$ that bound state becomes an open-channel ``shape resonance'', which is a quasi-bound  state above threshold trapped inside the centrifugal barrier of the p-wave potential. This leads to enhanced amplitude of the scattering wave function inside the barrier near the energy of the quasi-bound state, which manifests in the broad loss feature shown in Fig.~\ref{fig:BigPicture}(c). If we approximate the ``background'' scattering length for the fictitious $\ket{bb}$ s-wave channel to be that of the $^{40}$K$_2$ triplet potential, 169.2a$_0$, then $a/\bar{a} \approx 2.7$~\cite{Falke:2008dq}, and we can expect the $\ket{bb}$ channel to have such a shape resonance.  In fact, the quantum defect theory predicts a broad maximum in p-wave scattering amplitude inside the barrier around a collision energy of $E/h \approx$ 7\,MHz.  The resonance is broad and asymmetric, since this is an energy above the top of the p-wave barrier 5.8\,MHz, or 280\,$\mu$K.  The CC calculations demonstrate that such a p-wave shape resonance actually exists in this region, as indicated by the black dashed line in Fig.~\ref{fig:BigPicture}(c). The location and width of the shape resonance becomes a key parameter in the model developed in Sec.~\ref{sec:twochannel}. 

\subsection{Near-threshold molecular physics \label{ssec:NearThreshold}}

Since the ``last'' p-wave bound state in the $\ket{bb}$ channel is an above-threshold shape resonance, there are no other $\ket{bb}$ levels near threshold.  The actual last level with dominant $\ket{bb}$ character lies around 1.2\,GHz below threshold.  There is a cluster of p-wave components of mixed singlet-triplet character starting around $-120$\,MHz near $B=0$ and crossing threshold in the $200$\,G region.  The solid red line in  Fig.~\ref{fig:BigPicture}(c) shows the $M_\mathrm{tot}=-7$ level that interacts strongly with the $\ket{bb}$ and $\ket{ac}$ channels through short range spin-exchange to make the $M_L=0$ p-wave bound and quasi-bound levels studied in this experiment.  Figure \ref{fig:BigPicture}(d) shows the calculated energies of the $M_L=0$ and degenerate $M_L=\pm 1$ levels below threshold and their continuation above threshold as scattering resonances that broaden with increasing $E$ due to tunneling through the p-wave centrifugal barrier.  These levels are approximately 60\% singlet in character, and 80\% of their norm comes from a mixture of $\ket{aq}$ and $\ket{br}$ spin channels associated with one ground $9/2$ and one excited $7/2$ hyperfine component.

The green $M_\mathrm{tot}=-7$ level in Fig.~\ref{fig:BigPicture}(c) interacts only very weakly with the $\ket{bb}$ channel through spin-dipolar interactions.  It is a $M_L=-1$ level of dominant singlet character with 80\% of its norm coming from a mixture of the $\ket{ap}$ and $\ket{cr}$ channels.  Both bound levels in Fig.~\ref{fig:BigPicture}(c) have similar slopes at $B=120$\,G, $\delta\mu /h = 1.59$\,MHz$/$G for the lower and $\delta\mu /h=1.61$\,MHz$/$G for the upper.  The upper level is barely curved, having a slope of $1.51$\,MHz$/$G where it crosses threshold near 170.6\,G to make a very weak, narrow p-wave resonance (see App.~\ref{sec:NarrowFR}). By contrast, the bent lower level has a rapidly decreasing slope as it takes on more $\ket{bb}$ character in approaching threshold, having a value of $0.20$\,MHz$/$G at 198\,G just below its threshold crossing.  We estimate the approximate position of the ``bare'' (or ``undressed'') closed-channel bound state in the dashed red line of Fig.~\ref{fig:BigPicture}(c) by displacing the weakly interacting solid green line by 2.5\,G to higher $B$, which is the separation of the two states at 120\,G. This gives $B_\mathrm{n} \approx 173$\,G, roughly 26\,G below $B_0$. The parameter $\Delta_\mathrm{res} = \delta\mu (B_0 - B_\mathrm{n})$ is discussed further in Sec.~\ref{sec:SFB}. 

\begin{figure}[tb!]
\centering
\includegraphics[width=1.00\columnwidth]{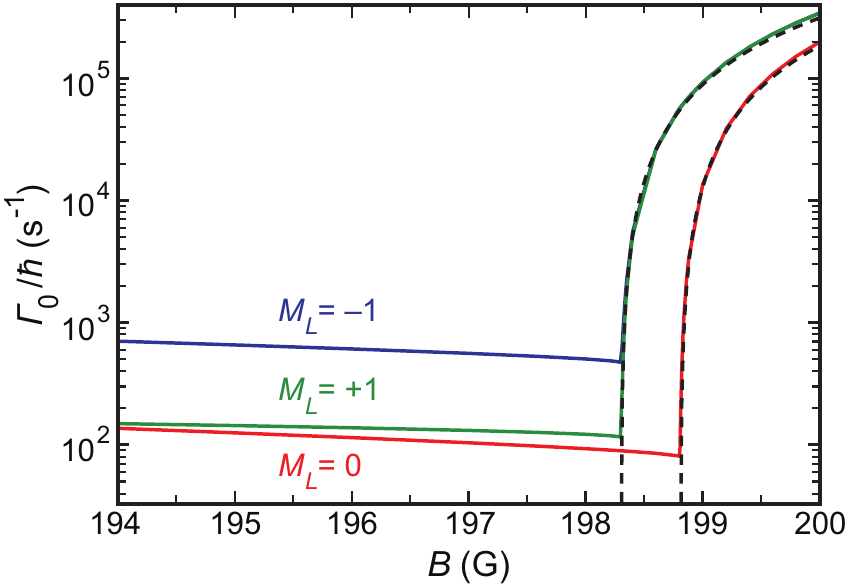}
\caption{  {\bf Feshbach dimer decay rates versus magnetic field $B$.}
Calculated decay widths of the $M_L=-1$, 0, and 1 levels in Fig.~\ref{fig:BigPicture}(d) from coupled-channels calculations that include p- and f-wave basis functions. The $M_L=\pm 1$ and 0 levels decay below the $\ket{bb}$ threshold to $\ket{ab}$ or $\ket{aa}$ exit channels through dipolar spin relaxation.  The decay rate of the $M_L=+1$ level below the $\ket{bb}$ threshold is smaller than the $M_L=-1$ level since it can decay only to centrifugally suppressed f-wave exit channels.  All levels show a rapid onset of decay by tunneling through the centrifugal barrier once the levels cross the $\ket{bb}$ threshold.  The level lifetimes are found by $\tau = 1/\Gamma_{0}$. Dashed lines show decay rates calculated from Eq.~\eqref{eq:Gamma0} using parameters from the bottom row of Tab.~\ref{tab:pwave}. 
\label{fig:DecayRates}}
\end{figure}

Figure \ref{fig:BigPicture}(c) shows the bending of the bound state of the broad Feshbach resonance as it approaches threshold from below.  While the scattering behavior in the continuum is complex, affected by the interaction of the red dashed ramping ``bare'' state with both the $\ket{bb}$ and $\ket{ac}$ channels indicated by strong $\ket{bb}$-to-$\ket{ac}$ loss mediated through the resonance, the qualitative picture in  Fig.~\ref{fig:BigPicture} suggests an ``avoided crossing'' between this ramping closed-channel state and the above threshold $\ket{bb}$ shape resonance in the $\ket{bb}$ channel.  The ``lower branch'' of the ``crossing'' connects the lower curving bound states Figs.~\ref{fig:BigPicture}(c) and \ref{fig:BigPicture}(d) with the shape resonance at high $B$, whereas the ``upper branch'' of the ``crossing'' distorts the shape resonance at low $B$ to join into an above-threshold p-wave Feshbach resonance of dominant singlet character extending to high $E$ and $B$. This resonance of the upper branch shows up prominently in the sloping broad (roughly $30$\,MHz wide) red contour of unitary loss where $1-|S_{bb,bb}|^2 \approx |S_{bb,ac}|^2 \approx 1$ in panel (c).  This quasi-bound feature, with a slope corresponding to a small absolute magnetic moment on the order of $-0.1$\,MHz/G in the 300\,G region, is the second p-wave level associated with a van der Waals potential with the singlet scattering length of 104a$_0$~\cite{Falke:2008dq} lying below the cluster of channels with separated atom spins of $9/2$ and $7/2$.

Since there is no strong loss from the $\ket{bb}$ to the $\ket{ac}$ channel below the $\ket{ac}$ threshold, which lies approximately 2\,MHz above $\ket{bb}$ in this region of $B$, Fig.~\ref{fig:BigPicture}(d) shows the elastic scattering probability in the near-$bb$-threshold region shown by the sum
\be \label{eq:sum}  
0 \leq \frac{1}{3}\! \sum_{M_{L}=-1}^{1} \frac{ \left|1-S_{bb,bb}(E,B;M_{L})\right|^2}{4} \leq 1 \,. \ee
In the limit of no loss (see Sec.~\ref{sec:twochannel}), $\frac14  |1 - S_{bb,bb}|^2 = \sin^2\delta_{bb}$. 
Just above threshold, the $M_L=0$ bound state and the two degenerate $M_L=\pm 1$ states emerge as quasi-bound levels that act as isolated normal p-wave resonances with well defined positions and widths, which we calculate using the algorithm in the molscat package~\cite{Hutson2019b,molscatNote,Frye2020}.  

The calculated widths are shown in Fig.~\ref{fig:DecayRates}. Below resonance, these calculations also make an updated prediction for the maximum lifetime of the p-wave Feshbach dimers: $\tau = \Gamma_0^{-1} \lesssim 2.1$\,ms for $M_L=-1$,  $\lesssim 12$\,ms for $M_L=0$, and $\lesssim 8.7$\,ms for $M_L=+1$. These lifetimes are roughly 30\% higher than the original predictions made in Ref.~\cite{Gaebler:2007}, primarily due to the inclusion of an effective spin coupling from Ref.~\cite{Xie2020}, which results in a smaller decay rate and consequently longer lifetime. The $M_L=0$ lifetime is also consistent with the experimental lower bound given in Ref.~\cite{Gaebler:2010}. 

\begin{figure}[tb!]
\centering
\includegraphics[width=0.99\columnwidth]{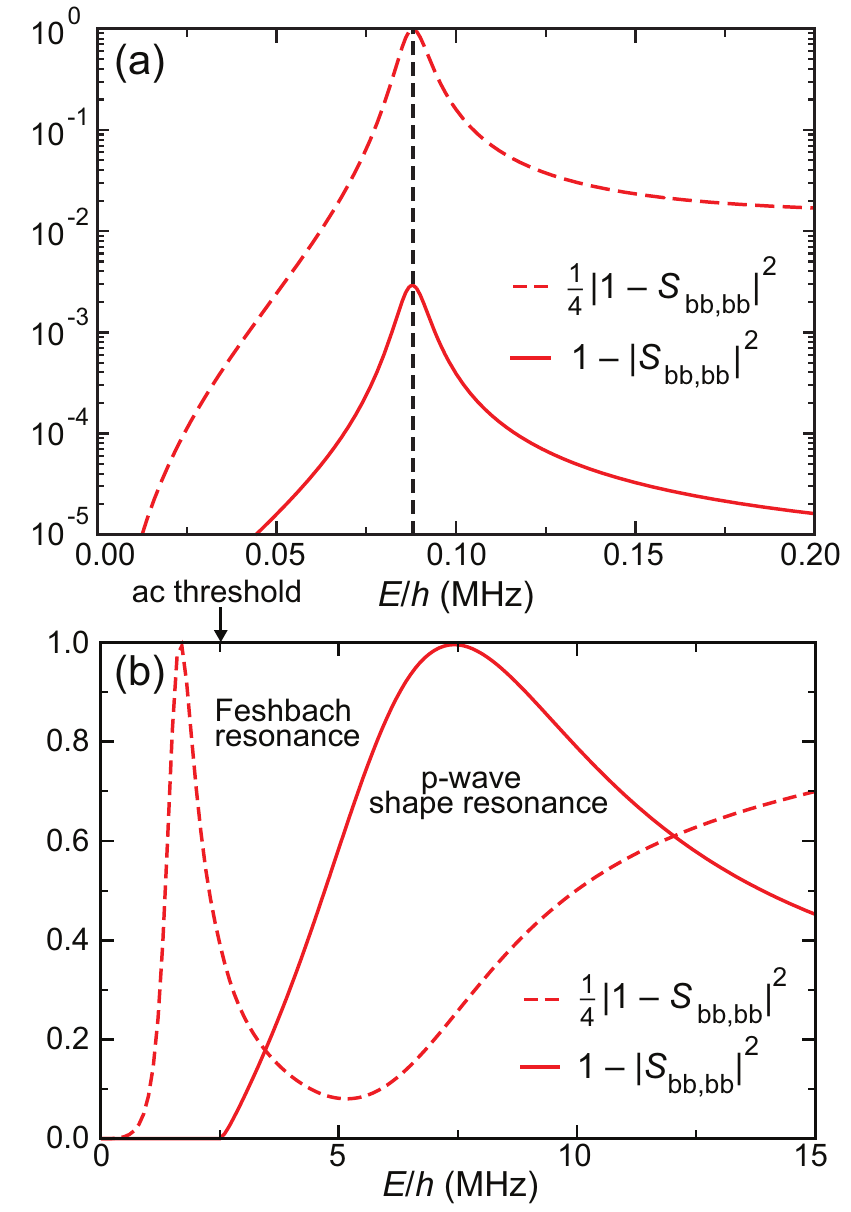}
\caption{{\bf p-wave scattering probabilities versus collision energy $E$.} 
(a) At 199.3\,G, the elastic scattering probability ($\frac14 |1-S_{bb,bb}|^2$, dashed line) and inelastic loss probability $(1-|S_{bb,bb}|^2$, solid line) on a log scale versus collision energy for the  the relatively narrow resonance with $M_\mathrm{tot}=-7$, $M_L=0$ at $E/h \approx 87.6$\,kHz with a width $\Gamma_0/h \approx 8.5$\,kHz.  The weak loss to the $\ket{ab}$ channel shows a peak at the same location as the unitary elastic scattering probability.  (b) Similar figure for the resonances at 210\,G.  While the Feshbach resonance near 1.5\,MHz with a width of 0.75\,MHz shows up prominently in elastic scattering probability, the p-wave shape resonance is prominent in the unitary inelastic loss from the $\ket{bb}$ channel to the $\ket{ac}$ channel.
\label{fig:LinePlots}}
\end{figure}

Figure~\ref{fig:LinePlots} shows elastic and inelastic scattering properties versus $E$ for cuts at constant $B$.  Both cuts show that the Feshbach resonances below the $\ket{ac}$ threshold feature prominently in the unitary peak in elastic scattering probability.  The shape resonance shows up prominently in inelastic loss, with a rapid onset versus energy when the $\ket{ac}$ channel opens near 2.5\,MHz in Fig.~\ref{fig:LinePlots}(b). The log plot in panel (a) shows that the weak dipolar loss from $\ket{bb}$ to the $\ket{ab}$ channel mirrors the Feshbach peak; a similar loss to $\ket{ab}$ would show up in a log plot of panel (b), but approximately one hundred times smaller in peak magnitude and much broader in resonance width due to the hundred-fold larger width at 210\,G as compared to 199.3\,G.  The larger width is due to much faster tunneling through the centrifugal barrier.

\section{Model for p-wave scattering \label{sec:twochannel}}

The coupled-channels analysis has allowed us to delineate an experimentally relevant regime of energies and fields in which $\ket{bb}$ is quite weakly coupled to loss channels, i.e.\ the limit in which the diagonal $|S_{bb,bb}| \to 1$. 
We now show that the scattering in this regime can be analyzed in terms of a simple model that retains only the shape resonance, the Feshbach resonances, and the dipole-dipole interaction. Written as a scattering $S$-matrix, 
\be \label{eq:modelS}
S = S_\mathrm{P} S_\mathrm{FB} S_\mathrm{dip} \, . \ee
In other words, this breaks the scattering phase into three contributions, $\delta_\mathrm{P}+\delta_\mathrm{FB}+\delta_\mathrm{dip}$. Our model is elastic, with real $\delta$ and unitary $S$ for each component.

In the following section, we derive simple analytic expressions that allow us to parameterize the field and energy dependence of the p-wave scattering phase. Although we fit the model parameters to the CC results for $^{40}$K, it should also provide a useful framework to understand elastic p-wave scattering in $^6$Li or any other cold gas.

\subsection{Open-channel shape resonance \label{sec:SP} }

As seen in Fig.~\ref{fig:LinePlots}(b), the p-wave shape resonance at 7\,MHz is asymmetric, in part because it lies near the top of the p-wave centrifugal barrier and in part because its strong coupling to the $\ket{ac}$ channel is truncated by the $\ket{ac}$ energetic threshold discussed in Sec.~\ref{sec:ccc}. Our model aims to capture only its effect on low-energy scattering, which is essential to understand the energy dependence of the scattering phase, as illustrated in Fig.~\ref{fig:PhaseVsEnergy}(a).

Inspired by the general form of Ref.~\cite{Ning:1948NingHuRepresentation}, we describe the open-channel resonance as 
\begin{align}
\label{eq:NingHu2poles}
S_\mathrm{P} = e^{-2i k r_0}\frac{(k-k_{\mathrm{s}}^*)(k+k_{\mathrm{s}})}{(k-k_{\mathrm{s}})(k+k_{\mathrm{s}}^*)},
\end{align}
where $k_{\mathrm{s}} = k_{\mathrm{R}} + i k_{\mathrm{I}}$ and $-k_s^*$ are the locations of the shape-resonance poles in the fourth and third quadrants of the complex $k$-plane ($k_\mathrm{I} < 0$). For any value of $k_s$, this form satisfies $S_\mathrm{P}^* S_\mathrm{P}=1$, $S_\mathrm{P}(-k) = S_\mathrm{P}^*(k)$. 
The background term $\exp(-2i k r_0)$ allows for an essential singularity at infinity and takes into account the effect of short range physics. We naturally expect $r_0$ to be on the order of $r_{\mathrm{vdW}}$. However, if we constrain $S_\mathrm{P}$ to follow the low-energy p-wave threshold law, then $r_0$ is constrained to be equal to
$2 \Im \{ 1/k_s \}=-2 k_\mathrm{I}/|k_s|^2$.
One can then match the $k\to0$ limit of $S_\mathrm{P}$ to find a scattering volume 
\be  \label{eq:VP}
V_\mathrm{P} = -r_0 |k_s|^{-2} \left(1 - {|r_0 k_s|^2}/{3} \right) \ee
and an effective range 
\be \label{eq:RP}
R_\mathrm{P} = \frac{ r_0 (1-|r_0 k_s|^2/3)^2}{1-|r_0 k_s|^2+|r_0 k_s|^4/5}
\ee 
such that $V_\mathrm{P} \approx -r_0 |k_s|^{-2}$ and $R_\mathrm{P} \approx r_0$ for $|k_I| \ll k_R$, as expected for a narrow low-energy resonance \cite{BEDAQUE2003159}. 
A similar form of the $S$-matrix is found in p-wave scattering from a square-well potential whose range is $r_0$ \cite{Nussenzveig}. Thus 
\begin{align}
\label{eq:Sppaper}
S_\mathrm{P} = \frac{e^{2k/k_s^*}}{e^{2k/k_s}}\frac{(k-k_{\mathrm{s}}^*)(k+k_{\mathrm{s}})}{(k-k_{\mathrm{s}})(k+k_{\mathrm{s}}^*)}.
\end{align}

\subsection{Feshbach Resonance \label{sec:SFB}}

The energy of ultracold collisions in the open channel is generally much smaller than the asymptotic energy of the closed channel to which the open channel is coupled. In addition, the bound states in the closed channel are typically spaced such that only a single bound state affects the open-channel state. Therefore, we can approximate the closed channel as a single bound state with bare energy $\epsilon_Q$ \cite{Kokkelmans:2014} that crosses the energy threshold at $B=B_{\mathrm{n}}$, and with a magnetic moment $\delta\mu$ relative to the open channel, such that $\epsilon_Q = \delta\mu (B-B_{\mathrm{n}})$.

Under this assumption, we can exploit the Feshbach formalism \cite{Feshbach:FESHBACH1958357,Feshbach:FESHBACH1962287,book:Feshbach} to write a two-channel model: 
\begin{align}
\label{eq:S2channelIntermediate}
S_\mathrm{FB} = 1-\frac{i\Gamma(E)}{E-\delta\mu (B-B_{\mathrm{n}})-A(E)} \, .
\end{align} 
Here $A(E)$ represents the complex-energy shift of the bare bound state $\ket{\phi_{\mathrm{b}}}$, 
\begin{align}
\label{eq:AEdefinition}
A(E) = \braket{\phi_\mathrm{b}|H_{\mathrm{QP}}\frac{1}{E^+-H_{\mathrm{PP}}}H_{\mathrm{PQ}}|\phi_\mathrm{b}},    
\end{align}
where $E^+ = E+i \epsilon$, with $\epsilon$ approaching zero from positive values. The labels of the coupling matrix strengths $H_{\mathrm{QP}} = H_{\mathrm{PQ}}^\dagger$ and the open-channel Hamiltonian $H_{\mathrm{PP}}$ refer to the open-channel subspace $\mathcal{P}$ and closed-channel subspace $\mathcal{Q}$.
Considering the multichannel nature of the system, we recognize that the subspace $\mathcal{Q}$ is comprised of multiple channel basis states whose composition changes as a function of the magnetic field due to the presence of the Zeeman Hamiltonian. Hence, the projection onto $\mathcal{Q}$ imparts a magnetic-field dependence to the coupling matrix strengths which would be absent in a true two-channel system. 

By inserting a complete set of eigenstates for the open-channel Hamiltonian, we can decompose Eq.~\eqref{eq:AEdefinition} as
\begin{align}
\label{eq:AEsplit}
A(E) = \Delta_{\mathrm{res}}(E) -\frac{i}{2}\Gamma(E),
\end{align}
where the real part $\Delta_{\mathrm{res}}(E)$ corresponds to the real energy shift of the bound state (estimated to be 26\,G in Sec.~\ref{sec:ccc}), and $\Gamma(E)$ adds a width to the Feshbach resonance. 
Since the propagator $(E-H_\mathrm{PP})^{-1}$ in Eq.~\eqref{eq:AEdefinition} shares its poles with the open-channel $S$-matrix $S_{\mathrm{P}}$ as introduced in Eq.~\eqref{eq:Sppaper}, we expect the complex-energy shift $A(E)$ to be well described in terms of a Mittag-Leffer series which runs over the poles of $S_\mathrm{P}$ \cite{Garcia-Calderon:GARCIACALDERON1976443,Romo:ROMO197861,Bang:BANG198089}, such that (suppressing factors of $\hbar^2/m$ for now)
\begin{align}
\label{eq:AEtwoPoles}
A(E) = &\frac{\braket{\phi_{\mathrm{Q}}|H_{\mathrm{QP}}|\Omega_{\mathrm{s}}}\braket{\Omega_{\mathrm{s}}^{\mathrm{D}}|H_{\mathrm{PQ}}|\phi_{\mathrm{Q}}}}{2k_s(k-k_s)} \notag \\
& - \frac{\braket{\phi_{\mathrm{Q}}|H_{\mathrm{QP}}|\Omega_{\mathrm{s}}^{\mathrm{D}}}\braket{\Omega_{\mathrm{s}}|H_{\mathrm{PQ}}|\phi_{\mathrm{Q}}}}{2k^*_s(k+k_s^*)},
\end{align}
where we have introduced the Gamow state $\ket{\Omega_{\mathrm{s}}}$, as well as its dual state $\ket{\Omega^{\mathrm{D}}_{\mathrm{s}}} \equiv \ket{\Omega_{\mathrm{s}}}^*$. Gamow states correspond to eigenstates of the open-channel Schr\"odinger equation with purely outgoing boundary conditions 
\cite{Gamow:Gamow1928}. Together with their dual states, Gamow states form a biorthogonal set, such that $\braket{\Omega_{\mathrm{s}}|\Omega^{\mathrm{D}}_{\mathrm{s'}}} = \delta_{\mathrm{s,s'}}$ \cite{Marcelis:2004hd,Romo:ROMO197861}. Since $\ket{\Omega_{\mathrm{s}}}$ is an eigenstate of $H_{\mathrm{PP}}$ with eigenvalue $E_s$ and $\ket{\Omega^{\mathrm{D}}_{\mathrm{s}}}$ is an eigenstate of $H_{\mathrm{PP}}^{\dagger}$ with eigenvalue $E_s^*$, these states follow the typical low-energy p-wave threshold behavior as implied by the Wigner threshold law \cite{taylor}. For energy-normalized states, this implies that $\Omega_{\mathrm{s}}\propto k_s^{3/2}$ and $\Omega^{\mathrm{D}}_{\mathrm{s}}\propto (k_s^{3/2})^*$, such that we can approximate the matrix-element $\braket{\phi_{\mathrm{Q}}|H_{\mathrm{QP}}|\Omega_{\mathrm{s}}}\braket{\Omega_{\mathrm{s}}^{\mathrm{D}}|H_{\mathrm{PQ}}|\phi_{\mathrm{Q}}}$ as ${g'} k_s^3 $, with momentum-independent coupling parameter $g'$. 
Substituting this approximation into Eq.~\eqref{eq:AEtwoPoles} and separating the real and imaginary contributions according to Eq.~\eqref{eq:AEsplit}, we find that 
\be \label{eq:DeltaResFB}
\Delta_{\mathrm{res}}(E) \approx g' \Re \left\{ \frac{E_s^{3/2}}{E-E_s} \right\} \ee
and
\be \label{eq:GammaFB}
\Gamma(E) \approx -2 g' \frac{E^{3/2} }{\left|E-E_s \right|^2} \Im \{ E_s \} \,. \ee
Since the pole of the shape resonance is independent of the external magnetic field, the full magnetic-field dependence of $\Delta_{\mathrm{res}}(E)$ and $\Gamma(E)$ is contained in the single parameter $g'$.  
Exploiting the distinct effects of the resonance shift and width, we retain only the lowest-order momentum dependence of the previous two expressions in the near-resonant regime, finding that $\Delta_{\mathrm{res}}(E) \approx \Delta_{\mathrm{res}}(0)+ \mathcal{O}(k^2)$ and $\Gamma(E) \approx g k^3 + \mathcal{O}(k^5)$, where $\Delta_{\mathrm{res}}(0)$ = $-g' k_{\mathrm{R}}$ and 
$g = 4 g'  k_{\mathrm{I}}k_{\mathrm{R}}  (k_{\mathrm{R}}^2+k_{\mathrm{I}}^2)^{-2}$ depend only on the complex momentum $k_{\mathrm{s}}$ and the parameter $g'$.
Keeping only the lowest-energy terms, Eq.~\eqref{eq:S2channelIntermediate} can be conveniently recast into the following simplified form around resonance
\begin{equation} \label{eq:S2channelFinal}
S_\mathrm{FB} \approx 1-\frac{i  g k^3}{E-\delta\mu(B-B_{\mathrm{n}})-\Delta_{\mathrm{res}}(0)+\frac{i}{2} g k^3} \,.
\end{equation}
We see that the momentum dependence of $S_\mathrm{FB}$ with this approximation follows the threshold scaling of Eq.~\eqref{eq:Seffrange}, with effective range
\be \label{eq:RFB} 
R_\mathrm{FB} = \frac{m}{\hbar^2} \frac{g}{2} \ee
and scattering volume
\be \label{eq:VFB}
V_\mathrm{FB} = -\frac{g/2}{\delta\mu(B-B_{\mathrm{n}})+\Delta_{\mathrm{res}}(0)}  \,.
\ee
At $B=B_0$, there is a resonance in $V_\mathrm{FB}$. In addition, since $g$ is B-field-dependent, Eq.~\eqref{eq:VFB} also has a background term. Keeping only the linear variation $g \approx g_0(1+\delta B/ \Delta_g)$, one recovers Eq.~\eqref{eq:scattvol}, $V_\mathrm{FB} = V_\mathrm{bg} (1 - \Delta/\delta B)$, with
\be \label{eq:VFB2}
V_\mathrm{bg}^{-1} \approx - \frac{m \Delta_g  \delta\mu}{\hbar^2 R_0} - \frac{|k_s|^2}{r_0} 
\quad \mbox{and} \quad \Delta \approx -\Delta_g \ee
where here $R_0$ is the value of $R_\mathrm{FB}$ at $B=B_0$.

When fitting the combined $S_\mathrm{P} S_\mathrm{FB}$ to coupled-channels data in Sec.~\ref{sec:parameterization}, we allow $\Delta_\mathrm{res}$ and $\Gamma$ to have independent variations with magnetic field. Relaxing this constraint can capture weak coupling to other channels with corrections to the positions 
of the poles of the $S$-matrix. We write this as 
\be \label{eq:SfittingFinal}
S_\mathrm{P} S_\mathrm{FB} =  \frac{e^{{2k}/{k_s^*}}(k-k_{\mathrm{s}}^*)(k+k_{\mathrm{s}})}{e^{{2k}/{k_s}}(k-k_{\mathrm{s}})(k+k_{\mathrm{s}}^*)} \frac{E-c - \frac{i}{2} g k^3}{E-c+\frac{i}{2} g k^3},
\ee
where $\delta\mu(B-B_{\mathrm{n}})+\Delta_{\mathrm{res}}(0)$ has been replaced by the fitting parameter $c(B)$, and $E = \hbar^2 k^2/m$. This is the p-wave analogue of the dual-resonant $S$-matrix 
presented in Ref.~\cite{Kokkelmans:2014} for s-waves. 

However, even this parameterized two-channel model will break down if we move sufficiently far away from resonance. Apart from the need to include the higher-order momentum dependencies of the resonance shift and width, the coupling to the $\ket{ac}$ channel as discussed in Sec.~\ref{sec:ccc} becomes increasingly important and we expect the need to update Eq.~\eqref{eq:SfittingFinal} to (at least) a three-channel model in order to accurately model the CC data.

\begin{figure}[tb!]
\centering
\includegraphics[width=0.99\columnwidth]{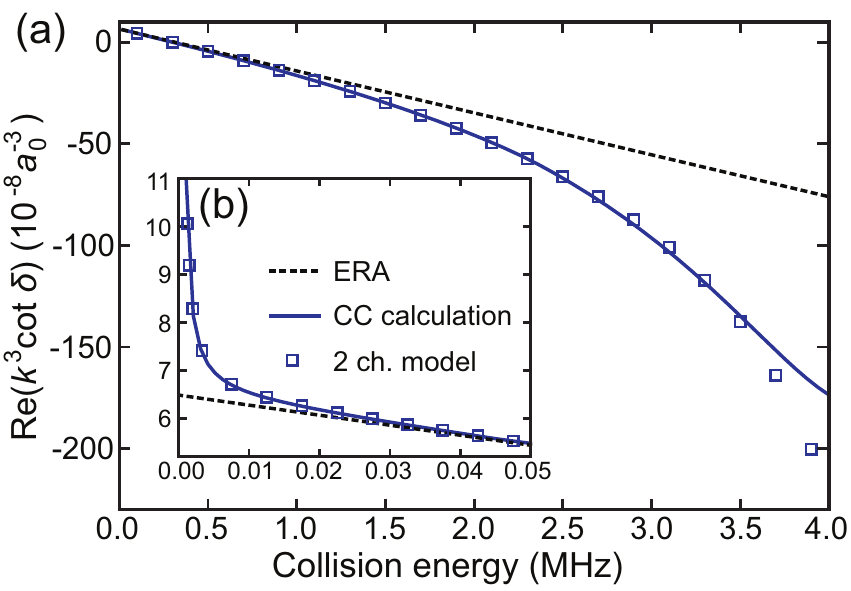}
\caption{{\bf Scattering phase at high and low energy} for $M_L=+1$ at 200\,G. 
The phase shift $\delta$ from CC calculations is plotted as the real part of $k^{3} \cot \delta$ (blue solid line) versus collisional energy $\hbar^2 k^2/m$.  (a) $k^{3} \cot \delta$ becomes nonlinear at higher collision energies, showing the necessity of including the shape resonance in the experimentally relevant regime. The best-fit model $S$ (blue squares) captures the deviation from the effective-range approximation (black dashed line) up to several MHz. (b) At low energy, $k^{3} \cot \delta$ has an apparent divergence due to the dipole-dipole interactions, as explained in the text. $S_\mathrm{dip}$ captures this effect with no fitting parameters. 
\label{fig:PhaseVsEnergy}}
\end{figure}

Figure \ref{fig:PhaseVsEnergy}(a) compares $S_\mathrm{P}S_\mathrm{FB}$ to CC calculations. 
The ERA captures only the linear variation of $k^3 \cot \delta$ with scattering energy, and requires a significant correction on the MHz scale \cite{Zhang:2010gc}. Using $k_s$ as a free parameter, our model captures the effect of the shape resonance out to several MHz. The results from fitting for a variety of magnetic fields near resonance are discussed further in Sec.~\ref{sec:parameterization}. 

\subsection{Dipole-dipole interaction \label{sec:dipole} }

A third contribution to the scattering phase is the long-range dipole-dipole interaction (DDI). As shown in Fig.~\ref{fig:PhaseVsEnergy}(b), the DDI dominates at low energy, causing a divergence in the scattering volume and invalidating the ERA \cite{OMalley:1961gh,Crubellier:2019kw}. However, the DDI phase shift in the open channel $\delta_{\mathrm{dip}}$ itself is small and is well treated by the Born approximation \footnote{Our approach treats the open-channel dipole-dipole interaction as a background phase shift. Reference \cite{PhysRevA.90.062714} describes how to treat the case where a distorted-wave correction to the Born approximation is required.}. In the limit $k \rightarrow 0$, 
\begin{align}
\label{eq:DeltaDipole1}
\delta^{L,M_L}_{\mathrm{dip}} \approx -\pi \braket{ijL M_L|V_3(\bm{r})|ijL M_L}, 
\end{align}
where we have introduced the dipole-dipole potential $V_3(\bm{r})$ acting on a channel state $\ket{ijL}$, with atoms in the internal states $\ket{\mathrm{i}}$ and $\ket{j}$ interacting with relative angular momentum $L$ and partial-wave projection $M_L$. 
Considering an external magnetic field oriented along the $z$-direction, we can express the dipole-dipole potential $V_3(\bm{r})$ as
\begin{align}
\label{eq:Vdipole}
V_3(\bm{r}) = -d^2 \frac{1-3\cos^2\theta}{r^3} = \frac{-2 d^2 Y_{20}}{r^3},
\end{align}
where $d$ is the magnetic dipole moment and $Y_{20}$ is a Racah-normalized spherical harmonic. Substituting Eq.~\eqref{eq:Vdipole} into Eq.~\eqref{eq:DeltaDipole1}, one finds
\begin{align}
\label{eq:DeltaDipole2}
\delta^{L,M_L}_{\mathrm{dip}} \approx \frac{\pi m d^2}{\hbar^2} \braket{L M_L|Y_{20}|L M_L} \int_0^{\infty}\frac{J_{L+1/2}(kr)^2}{r^2}dr,
\end{align}
where $J_{L+1/2}(kr)$ represents the Bessel function of the first kind. For $L=1$, we then obtain
\be \label{eq:Sdip}
\delta_\mathrm{dip}^{1,M_L} \approx -a^{1,M_L}_{\mathrm{dip}} k \ee
and $S_\mathrm{dip} = \exp (2 i \delta_\mathrm{dip}^{1,M_L}) $, where 
\be \label{eq:DeltaDipole3}
a_\mathrm{dip}^{1,0} = -\frac{m d^2}{5 \hbar^2} 
\quad \mathrm{and} \quad
a_\mathrm{dip}^{1,\pm 1} = \frac{m d^2}{10 \hbar^2} \,.
\ee
The same linear-in-$k$ scaling is found for all $L \geq 1$: $\delta_\mathrm{dip}^{L,M_L} = -(m_r C_3^{L,M_L} k/\hbar^2)/(L^2 + L)$ for scattering events with reduced mass $m_r$ from a potential $V=C_3^{L,M_L}/r^3$ \cite{MottMassey}. The characteristic length scale has been variously defined as $D_* = m d^2/\hbar^2$ \cite{Bortolotti:2006,Lahaye:2009kf}, or $a_\mathrm{d} = m_r C_3^{L,M_L}/\hbar^2$ \cite{Bohn2009}, which differ only by numerical factors calculated as in Eq.~\eqref{eq:DeltaDipole2}. 

In the $\ket{bb}$ channel, at $B=198.5$\,G, $d/\mu_B=-0.889$, such that $a_\mathrm{dip}^{1,0}=-0.153\,a_0$ and $a_\mathrm{dip}^{1,\pm 1}=0.077\,a_0$. The phase shift is small, and furthermore cancels when summed over all three $M_L$ channels. For these reasons, $\delta_\mathrm{dip}$ has been neglected in previous discussions of p-wave scattering of ultracold alkali atoms \cite{Gao:2009ek,Zhang:2010gc}, although it is this term that permits the evaporative cooling of spin-polarized Fermi gases in strongly dipolar species \cite{LevDFG,FerlainoDFG}. 

However in $^{40}$K, $\delta_\mathrm{dip}$ does not vanish because the $M_L=0$ is well separated from the $M_L=\pm1$ channels. Figure \ref{fig:PhaseVsEnergy}(b) shows that $\delta_\mathrm{dip}$ becomes the dominant phase shift at sufficiently low energy. In the low-$k$ limit, threshold scaling would give $\delta \to -V k^3$, which vanishes faster than $\delta_\mathrm{dip} \to -a_\mathrm{dip}^{1,M_L} k$. One sees that the threshold law is invalidated, and instead for $k^2 \ll |a_\mathrm{dip}^{1,M_L}/V|$, 
\be \cot{\delta} \to \frac{-1}{a_\mathrm{dip}^{1,M_L} k} + \frac{V}{(a_\mathrm{dip}^{1,M_L})^2} k + \mathcal{O}(k^3) \, .
\ee
With the usual definition $V \equiv -\lim_{k\to0} \tan\delta /k^3$, one would find a divergent $V(k) = a_\mathrm{dip}^{1,M_L}/k^2$. 

We see that $a_\mathrm{dip}^{1,M_L}/V$ determines a collision energy
\be E_\mathrm{dip}^{1,M_L} = -\frac{\hbar^2 a_\mathrm{dip}^{1,M_L}}{m V} \, , \ee 
at which there is a zero-crossing in the scattering phase when $E_\mathrm{dip}^{1,M_L} >0$. Since $|E_\mathrm{dip}^{1,M_L}| \ll E_F$, this appears as a low-energy divergence in $\cot\delta$ [seen in Fig.~\ref{fig:PhaseVsEnergy}(b)] for either $M_L$ state. For the background scattering lengths (see Sec.~\ref{sec:parameterization}), $E_\mathrm{dip}^{1,0}/h \approx -11$\,kHz and $E_\mathrm{dip}^{1,\pm1}/h \approx 6$\,kHz. Near the Feshbach resonance, this energy scale is even smaller: compared to the resonant energy given in Eq.~\eqref{eq:E0ERA}, $E_\mathrm{dip}^{1,M_L}/E_0 = a_\mathrm{dip}^{1,M_L}/R$, which is roughly $-0.003$ and $0.0015$ for $M_L=0$ and $|M_L|=1$, respectively.

In analyses of the elastic scattering near the p-wave Feshbach resonance, this weak dipolar effect causes a low-energy divergence in $\cot\delta$. We therefore fit $S_\mathrm{P} S_\mathrm{FB}$ to $S_\mathrm{dip}^{-1} S$ for the energy- and field-dependent $S$ from CC calculations. In general, we find that the Born approximation captures the DDI phase shift quantitatively below 10 microkelvin [see Fig.~\ref{fig:PhaseVsEnergy}(b)], with no fitting parameters. We can then discuss the low-energy limit of the reduced phase shift, $\delta - \delta_\mathrm{dip}$, in terms of a well defined scattering volume and effective range, recovering Eq.~\eqref{eq:effrange}. 

Dipole-dipole interactions can predominate in other scenarios. For instance $|E_\mathrm{dip}^{1,0}|/k_\mathrm{B} \gtrsim 2$\,$\mu$K across a ten-gauss range around the zero-crossing of $V$. Also, within molecular orbitals of the closed channel, dipolar interactions are responsible for the splitting of the Feshbach resonance into distinct $M_L=0$ and $|M_L|=1$ features \cite{Ticknor:2004}. 

\begin{figure*}[tb!]
\centering
\includegraphics[width=7in]{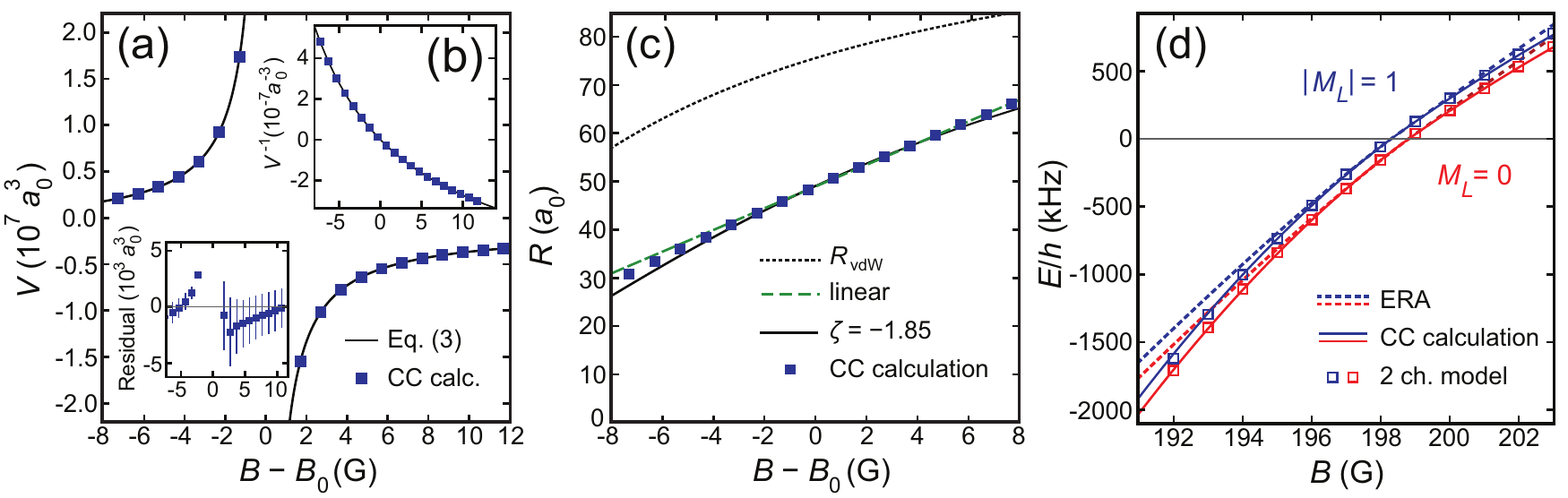}
\caption{{\bf Effective range parameters.} 
(a) Scattering volume $V$ and (b) inverse scattering volume $V^{-1}$ versus field-detuning from from Feshbach resonance. Blue squares are fits to the CC phase shift versus energy at each magnetic field; the black line is Eq.~\eqref{eq:scattvol} with values given in Tab.~\ref{tab:pwave}. The inset shows the  fit residuals. 
(c) The effective range $R$ versus field-detuning from resonance. Values determined through fits to CC phase shift versus energy (blue squares) are compared to the van der Waals limit of a broad resonance (black dashed and Eq.~\eqref{eq:Rbrd}), a single-parameter fit with the width parameter $\zeta$ (black solid line and Eq.~\eqref{eq:Rhyb}), and a linear fit (green dashed). Panels (a) and (b) show the $|M_L| = 1$ channel, but apart from a shift in $B_0$, the three channels have a similar behavior. 
(d) Dashed lines show bound-state poles from Eq.~\eqref{eq:kappaERA} below resonance, and the scattering resonance \eqref{eq:E0ERA} above resonance, using the parameterization of Eq.~\eqref{eq:scattvol} and Eq.~\eqref{eq:Rhyb}, with values given in Tab.~\ref{tab:pwave}. The points (open squares) are the poles of $S_\mathrm{FB}$. They are both compared to the CC binding energies (solid lines).
\label{fig:VandRvsB}}
\end{figure*}

\begin{table}[tb!]
 \centering
 \begin{tabular}{c|c}
 $\Re k_s$ & $0.572 \, /r_\mathrm{vdW}$ \\
 $\Im k_s$ & $-0.073 \, / r_\mathrm{vdW}$ \\
 $\Re E_s $ &  $0.295 \, \bar{E}$ \\
 $\Im E_s /2$ & $-0.038 \, \bar{E}$ \\
 $r_0$ & $0.44 \, r_\textrm{vdW}$\\
 $V_P$ & $-1.39 \, \bar{V}$ 
\end{tabular}
 \caption{{\bf Shape resonance parameterization.} The best-fit location of the poles of $S_\mathrm{P}$ are $k_s$ and $-k_s^*$, with values given above in terms of the van der Waals length $r_\mathrm{vdW} = \SI{65.0223}{\bohr}$. Also given for reference are the pole locations in the complex energy plane, $E_s$ and $E_s^*$, in terms of the van der Waals energy $\bar{E}/h = 23.375$\,MHz. The associated $r_0$ and $V_\mathrm{P}$ are also given in van der Waals units.
\label{tab:model}}
\end{table} 

\section{Field dependence of the scattering parameters \label{sec:parameterization}} 

Having understood both low- and high-energy dependence of the scattering phase, we can now fit our model to the scattering phase found by coupled-channels calculations. Using Eq.~\eqref{eq:SfittingFinal}, the variables $k_s$, $c$, and $g$ are fit to numerically generated $S_\mathrm{dip}^{-1} S$ across a range of collision energy and fields. We find that both $g$ and $c$ are approximately linear in magnetic field, while $k_s$ can remain at a fixed, field-independent value across the range of interest (see Tab.~\ref{tab:model}). This gives the collisional energy of the shape resonance as $\Re E_s/h = 6.88$\,MHz, and its effective width $-\frac12 \Im E_s/h=0.89$\,MHz.

The low-energy model parameters can be re-expressed in terms of the effective range parameters. 
When combining two p-wave $S$-matrices with scattering volumes $V_1$ and $V_2$, and effective ranges $R_1$ and $R_2$, the resulting scattering matrix has
\be \label{eq:Scombo}
V = V_1 + V_2 \quad \mbox{and} \quad \frac{1}{R} = \frac{V_1^2}{R_1 V^2} + \frac{V_2^2}{R_2 V^2} \, .
\ee
The scattering volume of $S_\mathrm{P} S_\mathrm{FB}$ is therefore
\begin{align}
\label{eq:Vshape}
V(B) = V_\mathrm{FB}(B) + V_\mathrm{P} \, ,
\end{align} 
where $V_\mathrm{FB}(B) = -\frac12 g(B)/c(B)$ and $V_\mathrm{P}$ is given by Eq.~\eqref{eq:VP}.
As discussed in Sec.~\ref{sec:SFB}, the first term creates the resonance of  Eq.~\eqref{eq:scattvol} and contributes a background due to the field dependence of $g$. The second term is independent of magnetic field, and here contributes $\sim 30\%$ of the total background scattering volume. 

\subsection{ Dispersive form of the scattering volume}

For the scattering volume near resonance, Eq.~\eqref{eq:scattvol} is a good approximation and we perform a straightforward fit to it, as presented in Figs.~\ref{fig:VandRvsB}(a) and \ref{fig:VandRvsB}(b). We find residuals at the $10^{-4}$ level, shown in the inset of Fig.~\ref{fig:VandRvsB}(a),
where the uncertainties correspond to the difference between fitting to the real and imaginary components of the $S$-matrix. We bound the residual field-dependent corrections to Eq.~\eqref{eq:scattvol} by looking at symmetric and antisymmetric combinations of $V(B)$ around $B_0$. For instance, $V_\mathrm{sym}(x)=V(B_0 + x)/2 + V(B_0 - x)/2$, which should give an $x$-independent $V_\mathrm{sym}(x)=V_\mathrm{bg}$ if Eq.~\eqref{eq:scattvol} is exact, or reveal corrections even in $\delta B$. Similarly, $V_\mathrm{as}(x)=V(B_0 + x)/2 - V(B_0 - x)/2$ should give an $x$-independent $V_\mathrm{as}(x) = V_\mathrm{bg} \Delta/x$. Here, we find a residual background slope of $\lesssim 10^{2}a_0^3/\mathrm{G}$ across the plotted range, below the $10^{-4}$ precision of the numerical determination. In sum, Eq.~\eqref{eq:scattvol} is an excellent parameterization of the CC-determined $V(B)$ near the p-wave Feshbach resonances of $^{40}$K.

The best-fit values of $B_0$, $V_\mathrm{bg}$, and $\Delta$ are given in Table~\ref{tab:pwave}. The accuracy of the resonance positions, $\pm$5\,mG (see Sec.~\ref{ssec:Fitting and systematics}), are improved by an order of magnitude when compared to the previous spectroscopic study \cite{Gaebler:2007}. 
The background scattering volume is primarily due to the open-channel van der Waals potential: $V_\mathrm{bg}$ estimated by using the triplet scattering length with Eq.~\eqref{eq:eta1} is $\approx -4.77 \bar V$, and here we find $V_\mathrm{bg} /\bar V \approx -4.84$.
The resonance width $\Delta$ predicts a zero-crossing of $V$ at $B_0-\Delta \approx 179$\,G. 
However, as discussed in Sec.~\ref{sec:dipole}, dipolar physics will dominate for ultracold collisions with small $V$, such that the ERA will no longer be valid. As pointed out by Refs.~\cite{Gao:2009ek,Zhang:2010gc,Hammer:2010hm}, the long-range nature of the van der Waals potential will also become increasingly relevant near the zero-crossing. 

\subsection{Parameterization of the effective range}

The effective range of $S_\mathrm{P} S_\mathrm{FB}$ is determined by the fit values of $k_S$, $c$, $g$, Eqs.~\eqref{eq:VP}, \eqref{eq:RP}, \eqref{eq:RFB}, and \eqref{eq:Scombo}, and plotted in Figure~\ref{fig:VandRvsB}(c) for $|M_L|=1$. 
A phenomenological linear fit to $R(B) = {R_0} (1 + \delta B/\Delta_{R})$ gives $\Delta_R = 21.1$\,G and $\Delta_R = 21.7$\,G for $|M_L|=0$ and $1$ respectively, and is shown as a green dashed line in Fig.~\ref{fig:VandRvsB}(c). As anticipated by Eq.~\eqref{eq:VFB2}, $\Delta_R \approx -\Delta$, since both are due primarily to the field dependence of the coupling $g$. 

\subsubsection{Narrow-resonance limit}

Insight into the relationship between $R(B)$ and $V(B)$ can be gained by considering a p-wave Feshbach resonance with $B$-field independent $R_{\mathrm{FB},0}={mg_0}/(2\hbar^2)$, such that $V_{\mathrm{FB},0} = -g_0/(2\,\delta\mu\,\delta B)$. The background scattering volume will then be solely due to the open channel, with $V_\mathrm{bg}$ and $R_\mathrm{bg}$. Using Eq.~\eqref{eq:Scombo}, one finds that the scattering volume, $V_{\mathrm{bg}}+V_{\mathrm{FB,0}}$, now matches the form of Eq.~\eqref{eq:scattvol} with $\Delta = g_0/(2\,\delta\mu\, V_{\mathrm{bg}})$. In the limit $V_\mathrm{bg}^2/R_\mathrm{bg} \ll V_\mathrm{FB,0}^2/R_\mathrm{FB,0}$, the effective range is controlled by $S_\mathrm{FB}$, and is 
\be \label{eq:RFB2}
R \approx  R_{\mathrm{FB},0} \left( 1 - \frac{\delta B}{\Delta}\right)^2 = 
R_{\mathrm{FB},0} \left(1- \frac{V_{\textrm{bg}}}{V(B)}\right)^{-2} \!\!\! .
\ee 
A similar form is found \cite{Zinner:2009gx,Blackley:2014bv} in the s-wave case for narrow resonances, where the effective range $r_e$ scales with scattering length $a$ as $(1 - {a_\mathrm{bg}}/{a(B)})^2$.

\subsubsection{Broad-resonance limit}

In the opposite limit, when the effective range is controlled by the open channel, one would expect $R$ to approach the van der Waals limit, $R_\mathrm{vdW}$, where \cite{Gao:2011ifa}
\be \label{eq:Rbrd}
R_\mathrm{vdW}(B) = R_\mathrm{max} \left( 1 + 2 \frac{\bar V }{V(B)} + 2 \frac{\bar V^2 }{V(B)^2 } \right)^{-1}\! . \ee
Here $R_\mathrm{max} \equiv 5 \bar V /(4  r_\mathrm{vdW}^2)$, and
$\bar V$ is given by Eq.~\eqref{eq:Vbar}. In the case of $^{40}$K, $R_\mathrm{max} \approx 1.162 r_\mathrm{vdW} \approx 76\,a_0$ is the maximum on-resonance value of the effective range allowed by causality \cite{Hammer:2009bd,Hammer:2010hm,Braaten:2012go} and the long-range van der Waals tail.

\subsubsection{General case}

For a resonance of mixed open- and closed-channel character, it has been found that a quadratic variation of effective range across the resonance is widely applicable to resonances of various widths and partial waves \cite{Blackley:2014bv}. The effective range on resonance is not expected to be universal, but can be taken as a fit parameter to experimental data or a CC calculation. One can define a dimensionless parameter $\zeta$ that characterizes the strength of the resonance by interpolating between Eqs.~\eqref{eq:Rbrd} and \eqref{eq:RFB2} in the broad $|\zeta| \gg 1$ and narrow $|\zeta| \ll 1$ limits respectively \cite{Gao:2011ifa,Werner:2012fs,Dong2016}, 
such that $R$ is given by 
\be \label{eq:Rhyb}
R^{-1}(B) =  - \frac{R_\mathrm{max}^{-1}}{\zeta}
\left( 1 - \frac{V_\mathrm{bg}}{V(B)} \right)^2 + R^{-1}_\mathrm{vdW}(B) \, , \ee
in which $-\zeta = R_0/(R_\mathrm{max}-R_0)$. 
Figure~\ref{fig:VandRvsB}(d) shows that fitting the single parameter $R_0$ can match both the resonant value of $R$ and its on-resonant linear slope $\equiv R_0/\Delta_R$, where from Eq.~\eqref{eq:Rhyb}, 
\be \Delta_R = -\Delta \left[2 - 2 \frac{R_0}{R_\mathrm{max}}\left(1+ \frac{\bar V}{V_\mathrm{bg}}\right) \right]^{-1}\,.\ee

The $\zeta$ values for the $M_L=0$ and $|M_L|= 1$ channels are $\{-1.90,-1.85\}$, respectively. Since $|\zeta| > 1$ the resonance can be considered broad. By comparison, one finds $\zeta \approx -4.1 \times 10^{-5}$ for the narrow resonance at 170\,G discussed in App.~\ref{sec:NarrowFR}, $\zeta \approx -0.3(1)$ for the most commonly used $^6$Li p-wave Feshbach resonance  \cite{Zhang:2004cy,Chevy:2005im,Schunck:2005cf,Fuchs:2008ka,Nakasuji:2013gw,Waseem:2017ep,Chang:2020,Marcum:2020}, and
$\zeta$ as large as $-560$ for p-wave resonances in bosonic $^{85}$Rb/$^{87}$Rb mixtures \cite{Dong2016}.

\begin{table}[t!]
 \centering
 \begin{tabular}{c|c|c|c|c}
  $B_0$~(G) & $V_{\rm bg}\, (a_0^3)$ & $\Delta$ (G) & $R_0\,(a_0)$ & Source \\ \hline 
    \multicolumn{5}{c}{} \\
\multicolumn{5}{c}{$M_L =0$} \\
     198.85 & -(101.6)$^3$ &  -21.95 & 47.2 & 
     \cite{Ticknor:2004} \\
     198.81(5) & & & & 
     \cite{Gaebler:2007} \\
     198.79(1) & & & & 
     \cite{Luciuk:2016gr} \\
     198.796(1)(5) & & & & 
     exp \\
     198.803 & -(108.0)$^3$ & -19.89 & 49.4 &  
     th \\ 
      \multicolumn{5}{c}{} \\
\multicolumn{5}{c}{$|M_L| =1$} \\
     198.37 & -(96.7)$^3$ & -24.99 & 46.2 & 
     \cite{Ticknor:2004} \\
     198.30(2) & & & & 
     \cite{Gaebler:2007} \\
     198.30(1) & & & & 
     \cite{Luciuk:2016gr}  \\
     198.293(1)(5) & & & & 
     exp \\
     198.300 & -(107.35)$^3$ & -19.54 & 48.9 & 
      th \\
\end{tabular}
 \caption{{\bf The $\bm{^{40}}$K p-wave Feshbach resonance parameters.} Comparison of experimental determinations of resonance location $B_0$, background scattering length $V_\text{bg}$, width $\Delta$, and on-resonant effective range $R_0$. The results from Sec.~\ref{ssec:Fitting and systematics} are listed as ``exp''; the fits discussed in Sec.~\ref{sec:parameterization} are listed as ``th''. When two uncertainties are listed, the first is statistical, and the second is systematic. 
}
 \label{tab:pwave}
\end{table} 

\subsection{ General properties of the S-matrix}

To be clear, even in the limit $|\zeta| \to \infty$, however, ultracold p-wave scattering is {\em energetically} narrow: $\Gamma_0/E_0 \to 0$ as $E_0 \to 0$, unlike s-wave scattering but similar to any $L \geq 1$ collision. Also unlike s-waves, one cannot neglect the effective range in the infinitely broad limit. 

The parameterizations of Eqs.~\eqref{eq:scattvol} and \eqref{eq:Rhyb} along with the pole location Eq.~\eqref{eq:kappaERA} below resonance or the scattering resonance energy Eq.~\eqref{eq:E0ERA} above resonance, give a closed-form expression for the p-wave dimer energy. Figure~\ref{fig:VandRvsB}(d) compares this energy directly to the CC energies, using the values given in Tab.~\ref{tab:pwave}. We see that the parameterization provides an accurate description of the dimer energies across several gauss. 

A more accurate determination of the dimer pole of the $S$-matrix can be found directly from $S_\mathrm{FB}$, instead of the ERA parameters. The pole below resonance can be found by solving $E_\mathrm{b}-c+i g k^3/2 = 0$ and taking the relevant root. Above resonance, because the Feshbach resonance is sharp across the range of interest, $E_0 \approx c$ is a good approximation of the scattering resonance. As shown in Fig.~\ref{fig:VandRvsB}(d), isolating this term in the two-channel model matches the ERA analytic result at very low energy ($E \ll E_s$, for which $S_\mathrm{P}$ can be taken to be unity) but improves agreement with the full CC calculation for energies $\gtrsim$ 0.5\,MHz, just as one would expect from Fig.~\ref{fig:PhaseVsEnergy}(a).

\section{Conclusion \label{sec:conclusion} }

We have used experimental and theoretical tools to probe the physics of ultracold p-wave collisions and near-threshold Feshbach dimer states.

A salient new phenomenon is the ``bending'' of the dimer energy, seen as a deviation from field-linearity in $E_\mathrm{b}$ (below resonance) and $E_0$ (above resonance). The clear observation of this nonlinearity is enabled by new analytic lineshape functions, resulting in excellent agreement with coupled-channels calculations, at a surprising milligauss scale. We explain, through both numerical and analytical models, that the origin of the curvature is the interplay of the ramping closed-channel state with the near-threshold shape resonance in the open channel. This provides a satisfying resolution to the discrepancy found in pioneering spectroscopy of the same resonance \cite{Gaebler:2007}. 

Coupled-channels calculations based on the full Hamiltonian proves to be an effective and accurate tool in understanding the collision physics of this dimer system. 
Calculations beyond the range of experimental measurements illustrate the complexity of collisions near the shape resonance, for which coupling between channels reach their unitary limit in this system. However, for ultracold (here, sub-MHz) collisions, cross-channel couplings become small and the physics is dominated by elastic scattering. In this regime we develop an analytic treatment based on a three-component $S$-matrix, which includes the effect of the p-wave Feshbach resonance, the open-channel p-wave shape resonance, and long-range dipole-dipole interactions. 

The dipolar phase shift causes the scattering volume to diverge at low energy due to its long-range character. 
This is strikingly different from the s-wave case. 
However since the absolute phase shift is small, it can be treated analytically, allowing us to isolate its contribution to the $S$-matrix and recover the conventional threshold scaling of scattering from the short-range component of the potential. 

Applied to the $^{40}$K p-wave resonances near 198.5\,G, we fit our model to the low-energy scattering phase generated by coupled-channels calculations. This allows an improved parametrization of the resonances, including location and effective range parameters ($V$ and $R$), as well as an effective shape-resonance location ($k_s$), and a parameterization of the Feshbach pole ($c$ and $g$). 

Several aspects of the collision physics can be understood approximately in terms of the open-channel van der Waals potential and quantum defect theory: the energy of the p-wave shape resonance, the background scattering volume \eqref{eq:eta1}, the relations \eqref{eq:DeltaResFB} and \eqref{eq:GammaFB} between the channel coupling and the complex-energy shift of the bare bound state, and the broad-resonance limit of the effective range \eqref{eq:Rbrd}. The comparison between the inferred $R$ at resonance and its maximum causality-limited value leads to a quantification of the ``width'' of the p-wave resonance, through Eq.~\eqref{eq:Rhyb}. 

Our work sets the stage for exploration in several ways. We have benchmarked coupled-channels calculations for two-body p-wave states in the continuum, building confidence in their extension to three-body states or strongly confined geometries. These calculations also make an updated prediction for the lifetime of the p-wave Feshbach dimers, relevant to feasibility of p-wave Fermi liquids \cite{Ding2019}. 
We have a better understanding of the limitations of the ERA, which is relevant to theoretical predictions based on this parameterization of the phase. 
 
\begin{acknowledgments} 
The authors would like to thank J.\ Gaebler and Y.\ Sagi for tabulations of their spectroscopy data, and J.\ Hudson and M.\ Frye for discussions regarding the use of the molscat package. We thank Jinlung Li for his assistance with the coupled-channels dipolar interactions and for stimulating discussions. This research is financially supported by AFOSR FA9550-19-1-7044, FA9550-19-1-0365, ARO W911NF-15-1-0603, NSERC, and the Netherlands Organisation for Scientific Research (NWO) under Grant No. 680-47-623. 

\end{acknowledgments}

\bibliography{pwave}

\clearpage
\newpage

\appendix
\section{Resonant association rate
\label{sec:TransitionRateappendix}}
In this section we outline the procedure to obtain the rate as introduced in Eq.~\eqref{eq:FermisGoldenRule} for a free-to-bound state transition in the RA protocol. We conclude with a discussion on how this framework can be generalized to the SFA protocol as well as to transitions to quasi-bound states.

In the RA protocol, where oscillation the longitudinal magnetic field results in coupling to a Feshbach dimer, we need to use (at least) a two-channel model in order to describe the incoming and outgoing states. The Hamiltonian of this two-channel model with open-channel subspace $\mathcal{P}$ and closed-channel subspace $\mathcal{Q}$ can be written as
\begin{align}
\label{eq:H2channel}
H_{\mathrm{2ch}} = \begin{pmatrix} H_{\mathrm{PP}} & H_{\mathrm{PQ}} \\ H_{\mathrm{QP}} & H_{\mathrm{QQ}} \end{pmatrix},     
\end{align}
where we have introduced the coupling matrix strengths $H_{\mathrm{QP}} = H_{\mathrm{PQ}}^\dagger$ and the open- and closed-channel Hamiltonians $H_{\mathrm{PP}}$ and $H_{\mathrm{QQ}}$ respectively. We can then define the following two-component wave function $\Psi(r,E)$
\begin{align} 
\label{eq:TwoChannelWaveFunction}
\Psi(r,E) = \phi(r,E) \ket{\mathcal{Q}} + \psi(r,E) \ket{\mathcal{P}},
\end{align}
where $\phi(r,E)$ and $\psi(r,E)$ represent the distance-dependent components of the total wave function in the two channels. 
The wave function $\Psi(r,E)$ of the full two-channel system is energy normalized, such that 
\begin{align} 
\int_{0}^{\infty} \Psi^+(E)^* \Psi^+(E')dr = \delta(E-E')
\end{align} 
and 
\begin{align} 
\int_{0}^{\infty} \Psi_{\mathrm{b}}(E_{\mathrm{b}}) \Psi_{\mathrm{b'}}(E_{\mathrm{b'}})\, dr = \delta_{b,b'},
\end{align}
with scattering wave functions $\Psi^+$ and bound state wave functions $\Psi_{\mathrm{b}}$. Using Eq. \eqref{eq:TwoChannelWaveFunction}, the orthogonality of the energy-normalized wave functions implies that 
\begin{align}
\label{eq:Orthogonality}
\int_{0}^{\infty} \phi^*_{\mathrm{b}}(r,E_{\mathrm{b}}) & \phi^+(r,E)\, dr \nonumber \\ &+ \int_{0}^{\infty} \psi^*_{\mathrm{b}}(r,E_{\mathrm{b}}) \psi^+(r,E)\, dr = 0.
\end{align} 
Here we note that, whereas the total wave function is orthogonal, the overlap between two open/closed channel wave function components with different energies is not necessarily zero. \par 
In general, the presence of an oscillating magnetic field $\mathbf{B}_{\mathrm{mod}}$ at frequency $\omega$ adds a contribution $H_{\mathrm{mod}}$ to the total Hamiltonian, where 
\begin{align}
\label{eq:Hosc}
H_{\mathrm{mod}} = -\bm{\mu} \cdot \mathbf{B}_{\mathrm{mod}}    
\end{align}
with magnetic moment $\bm{\mu}$.
As outlined in Sec.~\ref{ssec:Association Spectra}, the oscillating field in the RA protocol is aligned with the Feshbach field, such that $\mathbf{B}_{\mathrm{mod}} = B_{\mathrm{mod}} \, \hat{z} $ introduces a $\sigma_z$ coupling and hence drives $\Delta m_f = 0 $ transitions. In the dressed-state picture, a free-to-bound RA transition is between an initial state $\ket{\alpha,N}$ and a final state $\ket{\alpha',N+1}$, where $\alpha$ and $\alpha'$ represent the internal state of the two-atom system and where $N$ and $N+1$ indicate the number of quanta in the drive field. The Hamiltonian $H_{\mathrm{D}}$ can then be written as 
\begin{align}
\footnotesize{H_{\mathrm{D}} \to 
 \begin{pmatrix} \ddots \phantom{-------} &   \\
  \begin{bmatrix} H_{\mathrm{2ch}} \end{bmatrix}_N  & B_{\mathrm{mod}} \begin{bmatrix}  \mu_{\mathrm{P}} & 0 \\ 0 &   \mu_{\mathrm{Q}} \end{bmatrix}   \\[1em] 
B_{\mathrm{mod}} \begin{bmatrix} \mu_{\mathrm{P}} & 0 \\ 0 &  \mu_{\mathrm{Q}} \end{bmatrix} & \begin{bmatrix} H_{\mathrm{2ch}} \end{bmatrix}_{N} + \begin{bmatrix} \hbar \omega & 0 \\ 0 & \hbar \omega \end{bmatrix}  \\
& \phantom{---------} \ddots 
\end{pmatrix}}.  
\end{align}  

We can now apply Eq.~\eqref{eq:FermisGoldenRule} to the perturbing Hamiltonian $H_{\mathrm{mod}}$ of the RA protocol in order to find the transition rate from the initial free state $\ket{\alpha,N}$ to the final bound state $\ket{\alpha',N+1}$
\begin{align} 
\label{eq:TransitionRA}
\gamma &= \frac{2 \pi}{\hbar} \abs{\braket{\alpha, N+1 | H_{\mathrm{mod}} | \alpha' , N}}^2  \\ \notag 
&= \frac{2 \pi (\delta\mu \, B_{\mathrm{mod}})^2}{\hbar} \abs{\int_0^{\infty} \psi_{\mathrm{b}}^* (r,E_{\mathrm{b}})  \psi^+(r,E)}^2\, dr.
\end{align}
where $\delta\mu = \mu_{\mathrm{P}}-\mu_{\mathrm{Q}}$. 
Equation \eqref{eq:TransitionRA} implies the linear scaling of the transition rate with the Franck-Condon factor $F_{\mathrm{fi}}$ as introduced in Eq.~\eqref{eq:FCgeneral}, with initial state $\psi_{\mathrm{i}}(r) = \psi^+(r,E)$ and final state $\psi_{\mathrm{f}}(r) = \psi_{\mathrm{b}}(r,E_{\mathrm{b}})$ \par 
Following Refs.~\cite{Hanna2010,PhysRevA.92.022709}, the analysis presented in this section can be readily adapted to suit the SFA protocol. As stated in Sec.~\ref{ssec:Association Spectra}, the SFA protocol results in spin-flip transitions induced by the $\sigma^+$-polarization component of the rf field instead of the $\sigma_z$ coupling that is present in the RA protocol. As a result, the incoming state in the SFA protocol corresponds to $\ket{ab}$ and can be described by a single channel Hamiltonian. The outgoing state however still corresponds to $\ket{bb}$ and needs to be described (at least) by the two-channel Hamiltonian introduced in Eq.~\eqref{eq:H2channel}. In addition, one could use a similar approach to analyze a free-to-quasi-bound transition. Contrary to the free-to-bound transition, the consideration of the transfer towards a quasi-bound state with a higher energy than the incoming free state requires the absorption of a photon, such that the final state is $\ket{\alpha, N-1}$.

Due to the analogy between the SFA protocol and free to quasi-bound transitions on the one hand and the RA  transition analyzed in this section on the other hand, we can still generally write $\gamma \propto F_{\mathrm{fi}}$, where the  differences between the considered protocols and transitions result in different proportionality factors.

\section{Atom loss for transitions to bound and quasi-bound states
\label{sec:AtomLossappendix}}

Here we outline the assumed rate hierarchy underlying the perturbative atom-loss formulae in the main text. 
As mentioned in Sec.~\ref{ssec:lineshape}, we exploit the analogy between the PA protocol as discussed in Ref.~\cite{Jones:JonesReview2006} and the SFA and RA protocols in order to find a relation for the atom-loss. 
Reference \cite{Jones:JonesReview2006} used a field-dressed scattering formalism to obtain the two-body rate constant $K_\mathrm{loss}$ for atom loss driven on-resonance by an optical frequency laser.  This formalism is readily adapted to our case where the field-dressed states are coupled by a low-frequency oscillatory magnetic field instead of an optical frequency laser \cite{Kaufman2009,Hanna2010}.  We find that the number of atoms lost due to a near-resonant pulse of such radiation is
\begin{align}
\label{eq:NlossJones1}
\delta N_{\mathrm{loss}} =  a_N \int\! P(T,E) \frac{\hbar^2 \gamma_d \gamma(E)}{(E-E_k)^2+(\hbar \gamma_{\mathrm{tot}}(E)/2)^2} dE,
\end{align}
with a scaling factor $a_N$ that depends on experimental details (e.g., the time of the association pulse and the spatial distribution of the atoms in the trap) and a thermal factor $P(T,E)$. The necessity to average over a thermal distribution arises from the non-zero temperature of atoms in the the experimental set-up and will be discussed in detail in Sec.~\ref{ssec:Fitting and systematics}. We furthermore recognize the presence of three different rate constants $\gamma(E)$, $\gamma_d$ and $\gamma_{\mathrm{tot}}(E)$ in Eq.~\eqref{eq:NlossJones1}.
Here $\gamma_d$ represents the decay rate of the (quasi-)bound state due to all 1-, 2-, and 3-body decay processes that result in trap-loss, whereas $\gamma(E)$ represents the stimulated emission rate of (quasi-)bound-state atoms as presented by Eq.~\eqref{eq:FermisGoldenRule} back to the initial state $\ket{\mathrm{i}}$. The total decay rate $\gamma_{\mathrm{tot}}(E)= \gamma(E)+\gamma_d+\gamma_\mathrm{oth}$, where $\gamma_\mathrm{oth}$ represents any remaining part of the total decay rate that does not correspond to stimulated emission and that does not result in trap-loss.

As previously discussed in Sec.~\ref{ssec:Association Spectra}, we are working with experimental temperatures ranging from $0.2 \mu$K to $0.5 \mu$K. This means that the thermal factor $P(T,E)$ in Eq.~\eqref{eq:NlossJones1}  will have an energetic width in the order of $k_B T/h \sim 10$\,kHz. 
For free-to-bound transitions, this thermal width $k_B T$ is much larger than the decay width, such that $k_B T/\hbar \gg \gamma_{\mathrm{tot}}(E)$. Consequently, the lineshape in Eq.~\eqref{eq:NlossJones1} looks like a delta-function centered at $E_k$ with respect to the thermal spread in the energy, such that we can integrate over the Lorentzian and find \cite{Thorsheim:ThorsheimPA1987}
\begin{align}
\label{eq:NlossBound}
\delta N_{\mathrm{loss}} = A_N P(T,E_k) \frac{\gamma_d \gamma(E_k)}{\gamma_{\mathrm{tot}}(E_k)},    
\end{align}
where we have introduced the new scaling constant $A_N$ and where the energy is now fixed to $E_k$ defined by $E_\mathrm{b}$ and $\hbar \omega_\mathrm{osc}$. 
In both the SFA and RA protocol, we assume that the decay rate $\gamma_d$ of the (quasi-)bound state that leads to atom-loss is much larger than the rate of stimulated emission $\gamma(E)$  and the non-loss-related decay rate $\gamma_\mathrm{oth}$. This implies that the total decay rate can be approximated as $\gamma_{\mathrm{tot}} \approx \gamma_d$, such that Eq.~\eqref{eq:NlossBound} reduces to 
\begin{align}
\label{eq:NlossBound2}
\delta N_{\mathrm{loss}} = A_N P(T,E_k) \gamma(E_k),    
\end{align}
and the total number of atoms $N = N_{\mathrm{in}}-N_{\mathrm{loss}}$ can be calculated in accordance with Eq.~\eqref{eq:AtomNumberBound1}
as presented in the main text. \par 
Whereas we can still approximate $\gamma_{\mathrm{tot}} \approx \gamma_d$ in the case of free-to-quasi-bound transitions, we are no longer in the regime where the limit $k_B T/\hbar \gg \gamma_d$ applies. As analyzed in Sec.~\ref{ssec:NearThreshold}, the width of the bound state rapidly increases once the energy threshold is crossed. Consequently, the lineshape can no longer be regarded to be narrow with respect to the thermal distribution and we have to consider off-resonant transitions to contribute to the atom-loss. 
When $\gamma_d \approx \Gamma_0 > k_B T/\hbar$, the width term in the denominator exceeds the energy term in the low energy regime where the thermal factor is large, the variation with $E$ in the denominator is weak, and Eq.~\eqref{eq:NlossJones1} can be approximated by a form similar to Eq.~\eqref{eq:NlossBound2},
\begin{align}
\delta N_{\mathrm{loss}} = A_N' \int P(T,E) \gamma(E) dE,
\end{align}
where $A_N'$ once more represents an energy-independent constant and where the total number of atoms $N$ can now be calculated in accordance with Eq.~\eqref{eq:AtomNumberQuasiBound} as presented in the main text. 

\section{Franck-Condon factor for quasi-bound transitions \label{sec:FCappendix} }
The FC factor for transitions from incident wave vector $k$ to quasi-bound wave vector $\kq$ is found by evaluating Eq.~\eqref{eq:FCgeneral} with
\begin{align}
\psi_\mathrm{i} &= \sqrt{\frac{m}{\pi \hbar^2 k}} \left[\cos (\delta)\, \jhat_1(kr) + \sin(\delta) \, \hat n_1(kr) \right] \, \mbox{and} \nonumber \\
\psi_\mathrm{f} &= \sqrt{\frac{m}{\pi \hbar^2 \kq}} \left[\cos (\delta_\mathrm{q})\, \jhat_1(\kq r) + \sin(\delta_\mathrm{q}) \, \hat n_1(\kq r) \right]\,, 
\end{align}
where $\delta$ and $\delta_\mathrm{q}$ are the phase shifts of the free and quasi-bound states respectively. 
As outlined in Sec.~\ref{sssec:Free-to-bound transitions}, we neglect the short-range contribution to the overlap by cutting off all integrals at the length scale $r_c$. 
We find that 
\begin{align}
\label{eq:FCQBFull}
F_{\mathrm{fi}}(\kq) = &\frac{m^2}{\pi^2 \hbar^4 k \kq}\Big|\cos(\delta)\cos(\delta_{\mathrm{q}})\text{\RomanNumeralCaps{1}}+\sin(\delta)\sin(\delta_{\mathrm{q}})\text{\RomanNumeralCaps{2}} \notag \\
&+\cos(\delta)\sin(\delta_{\mathrm{q}})\text{\RomanNumeralCaps{3}}+\sin(\delta)\cos(\delta_{\mathrm{q}})\text{\RomanNumeralCaps{4}}\Big|^2,
\end{align}
where the integrals \RomanNumeralCaps{1}-\RomanNumeralCaps{4} are
\begin{align}
\text{\RomanNumeralCaps{1}} \equiv 
& \int_{r_c}^{\infty} \hat{\jmath}_1(\kq r) \jhat_1(k r) dr = \frac{\pi}{2}\delta(k-\kq)  \\
&+ \frac{k \sin(k r_c) \jhat_1(\kq r_c)-\kq\sin(\kq r_c) \jhat_1(k r_c)}{k^2-\kq^2} \nonumber \\
\text{\RomanNumeralCaps{2}} \equiv 
& \int_{r_c}^{\infty} \hat{n}_1(\kq r) \hat{n}_1(k r) dr = \frac{\pi}{2}\delta(k-\kq)  \\
&+ \frac{k \cos(k r_c) \hat{n}_1(\kq r_c)-\kq\cos(\kq r_c) \hat{n}_1(k r_c)}{k^2-\kq^2} \nonumber 
\end{align}
\begin{align}
\text{\RomanNumeralCaps{3}} \equiv& \int_{r_c}^{\infty} \jhat_1(k r) \hat{n}_1(\kq r) dr= \\
& \frac{k\sin(k r_c) \hat{n}_1(\kq r_c) - \kq \cos(\kq r_c) \jhat_1(k r_c)}{k^2-\kq^2} \notag \\
\text{\RomanNumeralCaps{4}} \equiv& \int_{r_c}^{\infty} \hat{n}_1(k r) \jhat_1(\kq r) dr= \\
& \frac{k \cos(k r_c) \jhat_1(\kq r_c) - \kq \sin(\kq r_c) \hat{n}_1(k r_c)}{k^2-\kq^2} \notag 
\end{align}
Here we have used Ref.~\cite{Cheng:1990Integrals} and the recursion relations $\jhat_{\ell+1}(z) = (1+2\ell)\jhat_\ell(z)-\jhat_{\ell-1}(z)$, and $\hat n_{\ell+1}(z) = (1+2\ell)\hat n_\ell(z)-\hat n_{\ell-1}(z)$. As outlined in Sec.~\ref{sssec:Free-to-quasi bound transitions}, Eq.~\eqref{eq:FCQBFull} reduces to the simplified expression Eq.~\eqref{eq:FCQBapprox} if the parameters in the SFA experiment satisfy Eq.~\eqref{eq:smallparams}. This reduces the computational time of the fitting to the experimental data significantly. 

\section{Narrow p-wave resonance at 170 G \label{sec:NarrowFR} } 

\begin{figure}[tb!]
\centering
\includegraphics[width=0.99\columnwidth]{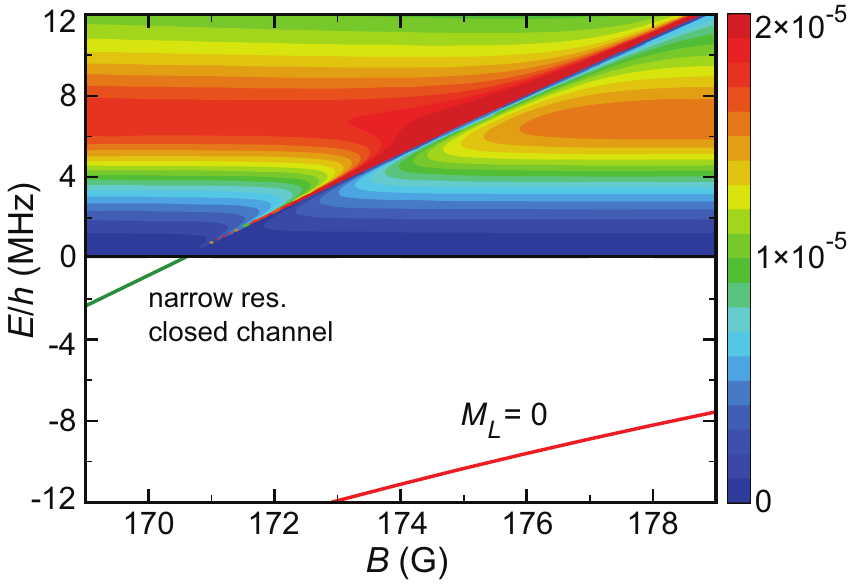}
\caption{  {\bf Bound and scattering properties for the narrow p-wave resonance with $M_\mathrm{tot}=-7$.}
The two bound states below the $\ket{bb}$ threshold are the same as in Fig.~\ref{fig:BigPicture}(c) with the same color coding; the energy zero is the separated atom energy of the $\ket{bb}$ channel. The contours above threshold show $|S_{bb,ab}|^2$ for weak decay from $\ket{bb}$ to $\ket{ab}$, with red and blue respectively indicating maximum and minimum magnitudes of $2 \times 10^{-5}$ and zero respectively.  The horizontal band centered around 7 MHz represents enhanced loss due to the $\ket{bb}$ channel shape resonance.
\label{fig:170Gregion}}
\end{figure}

Figure~\ref{fig:170Gregion} expands the region of the narrow resonance near 170\,G in Figure~\ref{fig:BigPicture}(c) to illustrate how the very weak resonance differs dramatically from the much stronger one near 198\,G studied in this paper.  The spin properties of the bound state making this narrow resonance are explained in Sec.~\ref{ssec:NearThreshold}.  The contour plot above threshold shows how this resonance shows up in the weak loss given by $|S_{bb,ab}|^2 \ll 1$.  We see that the quasi-bound state for $E>0$ does not bend with increasing $B$ as it crosses the broad horizontal shape-resonance region of enhanced background loss to $\ket{ab}$ centered around 7\,MHz.

Fitting the CC elastic scattering $S$-matrix element near 170\,G yields a small value of $V_\mathrm{bg} \Delta \approx$ 149 a$_0^3$ G, which with the value $\delta\mu/h=1.51$ MHz$/$G in Sec.~\ref{ssec:NearThreshold} gives a very small $\zeta=-4.1 \times 10^{-5}$. Here we have used the relation from Ref.~\cite{Gao:2011ifa} for an isolated resonance, 
$\zeta = -({m V_\mathrm{bg} \Delta \delta \mu})/({\hbar^2 R_\mathrm{max}}) = -0.8753 ({V_\mathrm{bg} \Delta \delta \mu})/({\bar{V}\bar{E}})$,
where the factor $0.8753 = \Gamma(\frac14)^4/(20\pi^2)$. This $\zeta$ is five orders of magnitude smaller than the value $\zeta \approx -2$ for the 198.5\,G resonance. 

\section{Discussion of the s-wave Feshbach resonance near 202 G \label{sec:swave}}

As an ancillary result, our updated $^{40}$K coupled-channels calculations provide some insight into the physics and parameterization of the commonly used s-wave Feshbach resonance, near 202\,G. First, we elucidate the relation of the s-wave resonance in the $\ket{ab}$ channel to the p-wave resonance in the $\ket{bb}$ channel that is the main topic of this paper. Then we comment on the discrepancy in parameterization of this resonance, and its possible resolution.

\subsection{Coupled-channels theory}

The same coupled-channels theory as described in Section~\ref{sec:ccc} is used to calculate the bound and scattering properties of the s-wave resonance in the $\ket{ab}$ channel, for which $M_\mathrm{tot}=-8$.  There are only two s-wave channels with this $M_\mathrm{tot}$ value, $\ket{ab}$ and $\ket{ar}$, which are strongly coupled by spin-exchange coupling. Adding d-waves to the calculation makes only mG-scale shifts in the calculated resonance positions.

\begin{figure}[tb!]
\centering
\includegraphics[width=0.99\columnwidth]{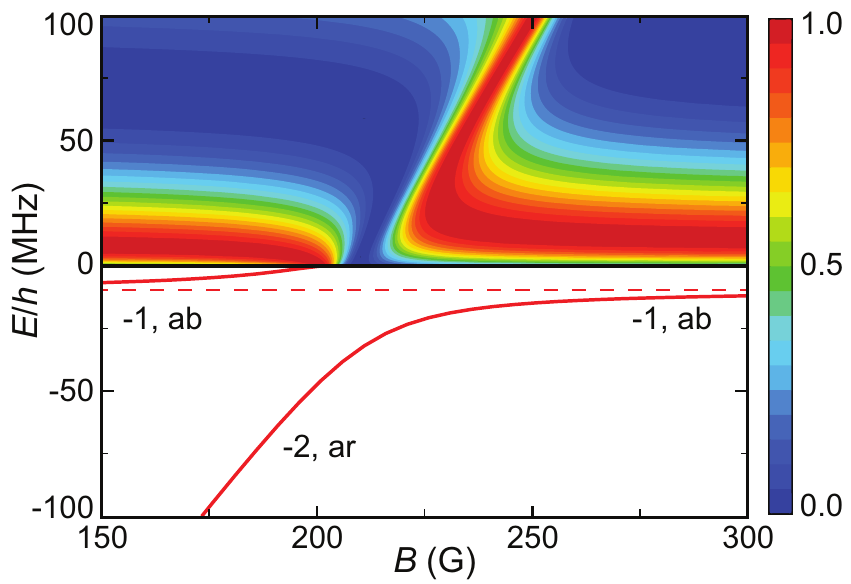}
\caption{  {\bf Bound and scattering properties for s-wave resonance with $M_\mathrm{tot}=-8$.}
The two bound states below threshold are the last bound state (labeled $-1,ab$) of the $\ket{ab}$ open s-wave channel and the next-to-last bound state (labeled $-2,ar$) of the closed $\ket{ar}$ s-wave channel; the energy zero is the separated atom energy of the $\ket{ab}$ channel. The dashed line shows the last bound state of a van der Waals potential with the triplet scattering length. The contours above threshold show $\sin^2\delta_{bb}(E,B)$, with red and blue respectively indicating unity and zero. 
\label{fig:abResonance}}
\end{figure}

Figure~\ref{fig:abResonance}, analogous to Fig.~\ref{fig:BigPicture}(c), shows the avoided crossing of the last bound state of the $|ab\rangle$ channel and the next-to-last bound state of the $|{ar}\rangle$ channel.  This avoided crossing ``pushes out'' the $\ket{ab}$ bound state at its threshold crossing of 202.110\,G.  The lower branch of the crossing ``returns'' to the $\ket{ab}$ level below threshold at high $B$ above the crossing.  The contour plot in the continuum shows that the ramping state reappears in the upper branch above threshold as a broad resonance feature sloping to the right.  The last $\ket{ab}$ bound state has entrance-channel character ($>99$\%) and is more than 90\% triplet over the range shown; for reference purposes, Fig~\ref{fig:abResonance} shows the energy of the last bound state of a van der Waals potential with the triplet scattering length.  The ramping state away from threshold has dominant $\ket{ar}$ character ($\approx$ 95\%) and is mainly singlet in character (more than 80\%), with a relatively low magnetic moment.  

One essential difference between the s-wave and p-wave cases is that the last bound state of the entrance channel ($\ket{ab}$ or $\ket{bb}$) is different: in the s-wave case, it is a real bound state below threshold; in the p-wave case, it is a shape resonance above threshold, as discussed in Sec.~\ref{sec:ccc}. Otherwise, the spin nature of the ramping states and the ``avoided crossing'' with the emergence of a broad resonance above threshold are similar in the s-wave and p-wave cases, with the threshold specifics being quite different because of the different nature of the threshold properties for the s-wave and p-wave cases. The universal quantum defect theory for a van der Waals potential predicts both the -9\,MHz s-wave “last” bound state and the +7\,MHz p-wave shape resonance connected with a van der Waals potential with the triplet scattering length.  In the 200~G region, both the $\ket{ab}$ and $\ket{bb}$ channels are predominately triplet in character, which is why the triplet scattering length makes a good background parameter for both channels (discussed in more detail in Sec.~\ref{sec:parameterization} and the next section), whereas the ramping state in both cases is predominantly singlet in character.

\begin{table}[tb!]
 \centering
 \begin{tabular}{c|c|c|c|c}
  $B_0$~(G)& $\Delta$~(G) & $B_\mathrm{zc}$~(G)& $a_\text{bg} (a_0)$& Source\\
  \hline
  \bf 201.5(14) & \bf 8.0(11) & 209.5 & \bf 174(7)  & \cite{Loftus:2002iz}\\
  \bf  & \bf 7.8(6) &  & 174 & \cite{Regal:2003go}\\
  \bf 202.10(7) & 7.8(6) & 209.9 & 174  &  \cite{Regal:2004kt}\\
  \bf 202.20(2) & \bf 7.04(10) & & 174 & \cite{Gaebler:2010}\\ 
  &  & \bf 209.6(1) &   &  \cite{Jordens:2010}\\ 
  & 7.0 & \bf 209.1(2) &   &  \cite{Schneider:2012ke}\\
   \bf 202.14(1) & \bf 6.70(3) & 208.84 & 169.7 &  \cite{Sagi2018}\\
    & & \bf 209.07(1) & & \cite{Smale:2019hd} \\
  202.15(2) & 6.910(3) & 209.06(2) & 166.978(2) & this work \\
 \end{tabular}
 \caption{{\bf The $\bm{^{40}}$K s-wave Feshbach resonance parameters.} Comparison of experimental determinations of resonance location $B_0$, width $\Delta$, zero-crossing field $B_\mathrm{zc}$, and the background scattering length $a_\text{bg}$. Directly measured quantities are indicated in bold face, while inferred or assumed values are in regular font.}
 \label{tab:swave}
\end{table}

\subsection{Discrepancy question}

The scattering length is typically parameterized as 
\be \label{eq:swaveaB}
a(B) = a_{\rm bg} \left( 1 - \frac{\Delta}{B - B_0} \right) \ee
where $a_{\rm bg}$ is the background scattering length, $B_0$ is the location of the Feshbach resonance, and $\Delta$ is the width of the resonance. This implies a zero-crossing of the scattering length of $B_\mathrm{zc}= B_0 + \Delta$. This resonance has been studied in numerous previous works  \cite{Loftus:2002iz,Regal:2004kt,Falke:2008dq,Gaebler:2010,Jordens:2010,Schneider:2012ke,Sagi2018,Smale:2019hd}. However, as precision has improved, a discrepancy has developed. As shown in Tab.~\ref{tab:swave}, $B_0$ determined by Refs.~\cite{Gaebler:2010} and \cite{Sagi2018} disagree by 60\,mG, or at least three sigma, and $B_\mathrm{zc}$ determined by Refs.~\cite{Sagi2018} and our previous work~\cite{Smale:2019hd} differ by 200\,mG, or at least six sigma.

By comparison, based on the potentials we use, our CC calculations find $B_0=202.110(1)$\,G and $B_\mathrm{zc}=209.018(1)$\,G. The value for $B_0$ is -30(10)\,mG from Ref.~\cite{Sagi2018}, and the value for $B_\mathrm{zc}$ is -50(10)\,mG from Ref.~\cite{Smale:2019hd}. This suggests that a +40(20)\,mG correction to the CC theory would give good agreement with both recent measurements. We show this updated parameter set as the last line of Tab.~\ref{tab:swave}: $B_0=202.15(2)$\,G and $B_\mathrm{zc}=209.06(2)$\,G. 

From CC data for $a(B)$, the background scattering length and quadratic correction terms can be extracted by plotting $a_\mathrm{sym}(x)=a(B_0 + x)/2 + a(B_0 - x)/2$ versus $x$, which should give an $x$-independent value of $a_\mathrm{bg}$ if Eq.~\eqref{eq:swaveaB} is correct. We find a small curvature ($\approx 3\times 10^{-5}a_0/\mathrm{G}^2$) leads to a $10^{-3}a_0$ variation in $a_\mathrm{sym}(x)$ across $\pm10$\,G. The best-fit $a_\text{bg}$ is 166.978(2)$a_0$. This value is roughly 2.5$a_0$ (2\%) lower than the triplet scattering length $a_\mathrm{T} \approx 169.5a_0$ from Ref.~\cite{Falke:2008dq}. 
As discussed in the previous section, the closed-channel state is primarily, but not purely, triplet. We estimate that roughly half of the discrepancy between the $\Delta = 6.70(3)$\,G found in Ref.~\cite{Sagi2018} and the value of  $\Delta$ we give here can be explained by their use of $a_\mathrm{bg}=a_\mathrm{T}$, since near-resonant spectroscopy constrains the product $a_\mathrm{bg} \Delta$. 

A similar analysis using $a_\mathrm{asym}(x)=a(B_0 + x)/2 - a(B_0 - x)/2$ should give an $x$-independent value of $x a_\mathrm{asym}(x) = a_\mathrm{bg} \Delta$ if Eq.~\eqref{eq:swaveaB} is accurate. We find a residual background slope of $\sim 1.2\times 10^{-2}a_0/\mathrm{G}$ across a $\pm10$\,G range. Including this term would decrease the best-fit $\Delta$ in Eq.~\eqref{eq:swaveaB} by 3\,mG. We use this systematic as an estimate of its uncertainty in the values in Tab.~\ref{tab:swave}. In sum, Eq.~\eqref{eq:swaveaB} is an excellent parameterization of the CC-determined $a(B)$ resonance across $\pm10$\,G, and consistent with the most recent and precise measurements of the Feshbach resonance when shifted by +40(20)\,mG. 

\end{document}